\documentclass[preprint]{aastex}
\def\pp{\parshape 2 0.0truecm 16.25truecm 2truecm 14.25truecm}
\newcommand{\be}{\begin{equation}}
\newcommand{\ee}{\end{equation}} 
\newcommand{\abig}{ {\cal A} } 
\newcommand{\lwig}{ {\widetilde L} } 
\newcommand{\uin}{ u_{in} } 
\newcommand{\vin}{ v_{\infty} } 
\newcommand{\vwig}{ {\widetilde v_{\infty}} } 
\newcommand{\uinss}{ u_{in{\rm c}} } 
\newcommand{\overdense}{ {\Lambda} } 
\newcommand{\muden}{ {\tilde \mu} }  
\def\lta{\,\raise 0.3 ex\hbox{$ < $}\kern -0.75 em
 \lower 0.7 ex\hbox{$\sim$}\,}
\def\gta{\,\raise 0.3 ex\hbox{$ > $}\kern -0.75 em
 \lower 0.7 ex\hbox{$\sim$}\,} 
\begin{document}

\title{Generalized Collapse Solutions with Nonzero Initial Velocities \\
for Star Formation in Molecular Cloud Cores} 

\author{Marco Fatuzzo,$^1$ Fred C. Adams,$^{2,3}$ and Philip C. Myers$^4$} 
 
\affil{$^1$Physics Department, Xavier University, Cincinnati, OH 45207} 

\affil{$^2$Michigan Center for Theoretical Physics, University of Michigan \\
Physics Department, Ann Arbor, MI 48109}  

\affil{$^3$Astronomy Department, University of Michigan, Ann Arbor, MI 48109}

\affil{$^4$Harvard Smithsonian Center for Astrophysics, 60 Garden Street, Cambridge, MA 02138} 

\email{fatuzzo@cerebro.cs.xu.edu, fca@umich.edu, pmyers@cfa.harvard.edu}

\begin{abstract} 

Motivated by recent observations that show starless molecular cloud
cores exhibit subsonic inward velocities, we revisit the collapse
problem for polytropic gaseous spheres. In particular, we provide a
generalized treatment of protostellar collapse in the spherical limit
and find semi-analytic (self-similar) solutions, corresponding
numerical solutions, and purely analytic calculations of the mass
infall rates (the three approaches are in good agreement). This study
focuses on collapse solutions that exhibit nonzero inward velocities
at large radii, as observed in molecular cloud cores, and extends
previous work in four ways: (1) The initial conditions allow nonzero
initial inward velocity.  (2) The starting states can exceed the
density of hydrostatic equilibrium so that the collapse itself can
provide the observed inward motions. (3) We consider different
equations of state, especially those that are softer than
isothermal. (4) We consider dynamic equations of state that are
different from the effective equation of state that produces the
initial density distribution.  This work determines the infall rates
over a wide range of parameter space, as characterized by four
variables: the initial inward velocity $\vin$, the overdensity
$\overdense$ of the initial state, the index $\Gamma$ of the static
equation of state, and the index $\gamma$ of the dynamic equation of
state. For the range of parameter space applicable to observed cores,
the resulting infall rate is about a factor of two larger than found
in previous theoretical studies (those with hydrostatic initial 
conditions and $\vin = 0$). 
 
\end{abstract}




\section{INTRODUCTION} 

Stars form through gravitational collapse and hence the dynamics of
the collapse flow is fundamental to understanding the star formation
process. Historically, theoretical studies of gravitational collapse
have considered initial conditions that correspond to hydrostatic
equilibrium (or near equilibrium).  In recent years, however,
observations have shown that some molecular cloud cores display
nonzero, inward velocities before they contain infrared sources. One
standard interpretation of this finding is that the cores must begin
their collapse with a head start, i.e., with a pre-existing inward
flow. The goal of this paper is to reexamine the gravitational collapse
problem for star formation with a focus on initial conditions that
include initial inward flows, i.e., the conditions observed in star
forming regions. Along the way, we generalize previous studies of
protostellar collapse to include a wider set of equations of state.

To isolate the effects of the inward flow, we return to the basic
spherical collapse problem for polytropic spheres, which are used as
idealized theoretical models (Shu 1977) of observed molecular cloud
cores (e.g., Myers \& Benson 1983, Benson \& Myers 1987).  This same
treatment can be applied to the collapse of subcondensations (denoted
as kernels -- see Myers 1998) within larger cores that form stellar
groups. In the simplest scenario, molecular cloud cores are thought to
evolve into centrally concentrated configurations through the process
of ambipolar diffusion (Mouschovias 1976; Shu 1983; Nakano 1984).
Since recent observations suggest that core formation takes place more
rapidly than predicted by standard models of ambipolar diffusion
(e.g., Jijina, Myers, \& Adams 1999), several modifications to this
picture have been suggested.  These generalizations include ambipolar
diffusion starting near the supercritical state (Ciolek \& Basu 2001),
turbulent fluctuations speeding up ambipolar diffusion rates (Fatuzzo
\& Adams 2002; Zweibel 2002), core formation through turbulent cooling
flows (Myers \& Lazarian 1998), and core (and star) formation through
dynamic, turbulent fragmentation (Li, Klessen, \& MacLow 2003; see
also Li 1999). In the extreme limit of some efficient fragmentation
scenarios, pre-collapse states (cores) never actually form at all. Our
present approach -- studying the collapse of cores that form with a
pre-existing inward flow -- is intermediate between the previous
picture of star formation starting from hydrostatic equilibrium (Shu,
Adams, \& Lizano 1987) and the alternate scenario of a completely
dynamic star formation process (MacLow \& Klessen 2003).  In any case,
molecular cloud cores are observed to form relatively rapidly and to
(often) exhibit subsonic inward motions in their starless state (Lee
et al. 1999, 2001). These configurations thus specify the initial
conditions for star formation and their subsequent collapse determines
the manner in which stars form.

The collapse of cloud cores can be studied using self-similar methods.
Indeed, self-similar collapse solutions (Shu 1977) are an important
cornerstone of the current theory of star formation (e.g., Shu et al.
1987). The original self-similar collapse calculations (Larson 1969ab;
Penston 1969ab; Shu 1977) considered an isothermal equation of state.
Since then, many generalizations of the collapse have been made. The
leading order effects of rotation have been studied, both for the
inner pressure free region (Ulrich 1976; Cassen \& Moosman 1981) and
for the entire core (Terebey, Shu, \& Cassen 1984).  Different
equations of state for the collapsing gas have been studied (Cheng
1977, 1978; McLaughlin \& Pudritz 1997). The leading order effects of
magnetic fields have also been included (Galli \& Shu 1993ab; Li \&
Shu 1996, 1997). More recently, the collapse of magnetized singular
isothermal toroids has been studied (Allen, Shu, \& Li 2003; Allen,
Li, \& Shu 2003); the latter study includes a calculation of collapse
with an imposed initial velocity (see their Figure 7), in the same
spirit as this investigation.  The effects of radiation pressure on
the collapse flow have been studied in both the spherical limit (e.g.,
Wolfire \& Cassinelli 1986, 1987) and the inner pressure free regime
for rotating collapse (Jijina \& Adams 1996).  Additional mathematical
generalizations of the solutions have also been considered (Hunter
1977; Whitworth \& Summers 1985).

The self-similar collapse solutions explored in this paper are
applicable for cores that exhibit a wide range of densities. Actual
molecular cloud cores have a finite central density $n_C$ that extends
over a given radial extent $r_C \sim a_S (2 \pi G \mu n_C)^{-1/2}$.
The core must also have an outer radial scale $r_{out}$ where the
declining density profile of the core matches onto the background of
the molecular cloud. If $r_M$ denotes the radius that encloses a
stellar mass, then self-similar solutions apply over the ordering of
radial scales $r_C \ll r_M \ll r_{out}$.  Previous numerical work
(Foster \& Chevalier 1993) indicates that the collapse of an
isothermal core approaches the expected self-similar form when the
core has $r_{out}/r_C > 20$. For typical values $M = 0.5$ $M_\odot$
and $a_S = 0.2$ km/s, this criterion implies a constraint on the
central number density $n_C > 3 \times 10^7$ cm$^{-3}$. In the context
of this paper, however, cores that have initial inward velocities can
more readily approach the self-similar collapse forms and these
constraints ($r_{out}/r_C > 20$; $n_C > 3 \times 10^7$ cm$^{-3}$) are
less severe.

Only a fraction of the observed starless cores have a sufficiently
extended range of densities (or, equivalently, large enough ratios
$r_{out}/r_C$) so that the collapse flows can be modeled to high
accuracy with the self-similar forms derived here. Many observed cores
have flat central regions (Tafalla et al. 2004), and tend to approach
the form $\rho \sim r^{-2}$ beyond a few thousand AU (Ward-Thompson,
Motte, \& Andr{\'e} 1999). In a recent survey (Bacmann et al. 2000), 3 of
the 24 observed cores also show evidence for a well defined edge
(corresponding to the scale $r_{out}$) so that these cores have small
ratios $r_{out}/r_C$. However, mapping results in Taurus are
consistent with a self-similar collapse picture and the envelopes in
Bok globules are also qualitatively consistent (Motte \& Andr{\'e} 2001). 
For cores with constant density centers, the infall rate will display 
more time variability than in the self-similar solutions: The infall 
rate will be very small at first, but quickly grows to become larger 
than that of the self-similar solution; once the constant density 
portion of the core has collapsed, the infall rate decreases toward 
the self-similar value (e.g., Foster \& Chevalier 1993; Myers 2004). 
Furthermore, even if actual cores only approach the self-similar
solutions asymptotically in time, this idealized formulation of the
collapse problem allows us to isolate the effects of various physical
inputs (especially nonzero initial velocities). 

The overall goal of this paper is thus to obtain a greater
understanding of the collapse of molecular cloud cores.  This work
includes four generalizations of previous studies of protostellar
collapse, but the most important astronomical issue is to include the
effects of pre-existing inward flows, which have now been observed (as
outlined above). We can include an initial inward velocity as part of
the initial conditions (i.e., we let $v \ne 0$ at large radii).  In a
similar vein, we can consider initial states that are overdense and
thus exceed the density required for hydrostatic equilibrium. These
two complications are related in that both can explain the observed
inward flows and both lead to larger mass infall rates, but they are
conceptually distinct.  In the first case, the inward flows are part
of the initial conditions and must be produced by whatever physical
processes form the core. In the case of overdense initial states,
however, the inward flows are part of the collapse. Specifically, in
the latter case, the core begins with zero initial velocity throughout
its structure, but the overdense core soon develops small, but
nonzero, inward speeds at large radii (as observed).

In addition, we consider equations of state that are softer than
isothermal. This generalization is motivated by observations of
non-thermal linewidths $\Delta v$ in molecular cloud cores, where one
finds correlations with density of the form $(\Delta v)^2 \sim
\rho^{-\beta}$, with $\beta > 0$ (e.g., Larson 1985, Jijina et
al. 1999).  If one interprets the linewidth $\Delta v$ as the
effective transport speed in the medium, then the corresponding
effective equation of state has the form $P \sim \rho^\Gamma$, where
$\Gamma = {1-\beta}$ and hence $0 < \Gamma \le 1$.  In the limit
$\Gamma \to 0$, $P \to \log \rho$; this limiting case of a sequence of
possible equations of state is often called a ``logatrope'' (beginning
with Lizano \& Shu 1989). Here we consider a range of effective
equations of state, from the isothermal limit $\Gamma = 1$ down to the
logatropic limit $\Gamma \to 0$ (collapse solutions in the logatropic
limit have been studied previously by McLaughlin \& Pudritz 1997). 
Notice that the general equations we work with below reduce to the
logatropic case in the limit $\Gamma \to 0$ (but care must be
exercised in taking the limit).

As another generalization, we consider the case in which the dynamic
equation of state for the collapsing gas is different than the
effective (static) equation of state that produces the initial
equilibrium configuration. Here, the static equation of state refers
to the pressure law that enforces the initial (pre-collapse)
configuration for the core. Note that we consider only barotropic
equations of state throughout this paper.  The dynamic equation of
state (as set by $\gamma$) refers to the pressure law that describes
how the thermodynamic variables of the gas change as the material is
compressed during collapse; this process is governed by the entropy
evolution equation (see \S 2.1). Equations of state that are softer
than isothermal are motivated by the observed linewidth dependence on
density, and the inferred density profiles, in molecular cloud
regions. These considerations apply primarily to the static equation
of state as defined here (for applications of such soft, static
equations of state, see, e.g., Curry \& McKee 2000, Curry \& Stahler
2001, Spaans \& Silk 2000). The dynamic equation of state, as set by
the index $\gamma$ that appears in equation (\ref{eq:entropyone}),
need not be the same as the static equation of state. In other words,
the physics that determines the density profiles of the pre-collapse
states can be different from the physics that governs the
thermodynamics of the collapse flow. We thus allow $\gamma \ne
\Gamma$. This possibility is also motivated by theoretical MHD
simulations (Vazquez-Semadeni, Canto, \& Lizano 1998) which show that
the turbulent velocity dispersion increases with mean density as the
gas collapses, in contradiction to the expected behavior for a soft
equation of state; as a result, logatropic and similarly soft
equations of state may provide an inadequate description of the
dynamical processes occurring within a cloud.

The mass infall rate $\dot M$ is one of the most important physical
quantities in the star formation problem and this paper focuses on
calculating $\dot M$ for a range of conditions.  Collapse flows are
often self-similar and have no characteristic mass scale. Instead, the
collapse flow feeds material onto the central star/disk system at a
well-defined mass infall rate $\dot M$. In these flows, the infalling
material always approaches free-fall conditions on the inside (in the
limit $r \to 0$) and the reduced mass $m_0$ determines the size of the
infall rate (as shown below).  The infall rate $\dot M$ determines, in
part, the total system luminosity and the total column density of the
infalling envelope.  These quantities, in turn, largely account for
the spectral appearance of protostellar objects (e.g., Adams, Lada, \&
Shu 1987; Adams 1990). For protostellar objects, most of the
luminosity is derived from material falling through the gravitational
potential well of the star. Although the circumstellar disk stores
some of the energy in rotational motion, the system luminosity is
(usually) a substantial fraction of the total available luminosity 
\be 
L_0 \equiv {G M {\dot M} \over R_\ast } \, , 
\label{eq:luminosity} 
\ee 
where $M$ is the total mass of the system, ${\dot M}$ is the mass
infall rate, and $R_\ast$ is the stellar radius. The stellar radius,
which helps determine the depth of the potential well, is itself a
function of the mass infall rate (Stahler, Shu, \& Taam 1980).

This paper is organized as follows. In \S 2, we present a general
formulation of the collapse problem, including both the similarity
transformation for self-similar solutions and a complementary
numerical treatment. We present general analytic results in \S 3,
including a discussion of how the ``outer'' spherically symmetric
solutions of this paper match onto collapse solutions in the inner
regime where the effects of rotation are important.  In \S 4, we
consider collapse solutions for which the static equation of state is
isothermal (so that the initial state $\rho \sim r^{-2}$), but
generalize previous work to include nonzero initial flow speeds $(\vin
> 0)$, overdense initial configurations $(\overdense > 1)$, and
dynamic equations of state that are not isothermal ($\Gamma = 1$,
$\gamma \ne 1$).  In \S 5 we consider collapse solutions from
non-isothermal initial conditions, i.e., starting states for which the
static $\Gamma \ne 1$. In \S 6, we conclude with a summary and
discussion of our results.

\section{FORMULATION OF THE COLLAPSE PROBLEM}  

In this section we formulate the collapse problem for the general case
of a collapsing spherical cloud of gas. The collapse solution is
obtained by self-similar methods (\S 2.2) and by a numerical approach
(\S 2.3). In both cases, we allow the initial states to be out of
exact hydrostatic equilibrium by having initial inward velocities
and/or by being overdense. We also allow the dynamic equation of state
to be different from the static equation of state that determines the
initial density distribution.

\subsection{Basic Governing Equations}  
 
The basic equations governing the fluid are given below.  We want to
consider the spherical collapse problem, so that the equations of
motion reduce to a simpler form.  Conservation of mass can be
expressed in terms of mass shells  
\be
{\partial M \over \partial t} + u {\partial M \over \partial r} = 0  
\qquad {\rm and} \qquad 
{\partial M \over \partial r} = 4 \pi r^2 \rho , 
\ee
where $M(r,t)$ is the total mass within a given radius $r$ at a time
$t$. The above equations are equivalent to the usual equation of 
continuity, 
\be
{\partial \rho \over \partial t} + {1 \over r^2}
{\partial \over \partial r} (r^2 \rho u) = 0 \, ,  
\label{eq:fullcont} 
\ee
where $\rho$ is the density and $u$ is the radial component of 
the velocity. The force equation can be written in the form 
\be
{\partial u \over \partial t} + u
{\partial u \over \partial r} = - {1 \over \rho} 
{\partial P \over \partial r} - {G M \over r^2} \, ,  
\ee
where $P$ is the pressure.  Finally, the conservation of entropy 
equation can be written in the form 
\be
\bigl( {\partial \over \partial t} + u {\partial \over \partial r} 
\bigr) \bigl[ \log (P/\rho^\gamma) \bigr] \, = 0 \, , 
\label{eq:entropyone} 
\ee
where $\gamma$ is the index of the dynamic equation of state.
Adopting equation (\ref{eq:entropyone}) is not the same as assuming an
equation of state of the form $P = K \rho^\gamma$. Equation
(\ref{eq:entropyone}) describes how a given parcel of gas changes its
thermodynamic variables along a streamline, whereas the equation of
state ($P = K \rho^\gamma$) implies a global constraint on those
variables.

\subsection{The Similarity Transformation}  

In this section, we find a similarity transformation for the collapse
problem formulated above.  Mathematically, this problem is represented
by coupled partial differential equations in the variables time $t$
and radial position $r$. Using the similarity transformation, we can
reduce this problem to a set of ordinary differential equations in a
new similarity variable $x$ which we define below. In particular, we
look for a similarity transformation of the general form 
$$
x = A t^a r \, , \qquad 
\rho = B t^b \alpha(x) \, , \qquad 
M = C t^c m(x) \, , 
$$
\be
u = D t^d v(x) \, , \qquad {\rm and} \qquad 
P = E t^e p(x) \, . 
\ee
Here, both the coefficients ($A$, $B$, $C$, $D$, $E$) and the indices
$(a, b, c, d, e)$ are constants.  The reduced fluid fields ($\alpha$,
$m$, $v$, $p$) are dimensionless functions of the (single)
dimensionless similarity variable $x$. The time benchmark $t=0$ 
corresponds to the instant of the onset of collapse when the mass of
the central ``object'' is zero, i.e., $M(0,t)$ = 0. 

The general similarity transformation calculation leads 
to four equations to specify the five indices $a,b,c,d,e$.  
We leave the constant $a$ arbitrary for the moment and write 
the rest of the variables in terms of its value, i.e., 
$$
a=a \, , \qquad b = - 2 \, , \qquad c = - (3a + 2) \, , 
$$
\be
d= - (a + 1) \, , \qquad {\rm and} \qquad 
e = - 2 (a + 2) \, . \ee
Similarly, for the coefficients we obtain
$$
A = A \, , \qquad 
B = (4 \pi G)^{-1} \, , \qquad 
C = (A^3 G)^{-1} \, , 
$$
\be
D = A^{-1} \, , \qquad {\rm and} \qquad 
E = (4 \pi G A^2)^{-1} \, . 
\ee
Keep in mind that $G$ is the gravitational constant. 
We thus obtain reduced equations of motion in the form 
\be(ax + v) {dm \over dx} = (3a + 2)  m \, , 
\label{eq:contmass} 
\ee 
\be
{dm \over dx} = x^2 \alpha \, , 
\label{eq:contmass2} 
\ee 
\be
(ax + v) {dv \over dx} + {1 \over \alpha} {dp \over dx} 
= - {m \over x^2} + (a+1) v \, , 
\label{eq:force} 
\ee 
\be
(ax + v) {d \over dx} \log [ p/\alpha^\gamma ] = 
2 (2 + a - \gamma) \, , 
\label{eq:entropy} 
\ee 
\be
(ax + v) {1 \over \alpha} {d \alpha \over dx} + 
{dv \over dx} = 2 (1 - v/x) \, . 
\label{eq:contrho} 
\ee
This similarity transformation is not unique -- one can always rescale
the coefficients $\{ A, B, C, D, E \}$ by a set of dimensionless
numbers and obtain new equations of motion with different numerical
coefficients.

Notice that we can immediately combine the first two equations of
motion to obtain an expression for the reduced mass $m(x)$, i.e., 
\be
m = {(ax + v) \over (3a + 2) } x^2 \alpha \, . 
\label{eq:massint} 
\ee
Next we define a constant $q$ according to  
\be
q \equiv {2 \over 3a + 2} (2 + a - \gamma) . 
\label{eq:qdef} 
\ee
If we then take $q$ times equation (\ref{eq:contmass}) and subtract 
it from equation (\ref{eq:entropy}), we can integrate the resulting 
equation to obtain an expression for the reduced pressure
$p(x)$, i.e.,  
\be
p = {\cal C}_0 \alpha^{q + \gamma} \, (x^3 + v x^2/a)^q \, 
= \, {\cal C}_1 \, \alpha^{\gamma} \, m^q \, , 
\label{eq:psolution} 
\ee
where ${\cal C}_0$ and ${\cal C}_1$ are constants which depend on the
initial density distribution of the core.  Given these solutions for
the reduced pressure $p(x)$ and reduced mass $m(x)$, equations
(\ref{eq:force}) and (\ref{eq:contrho}) are the relevant equations of
motion to determine the remaining unknown functions $\alpha(x)$ and
$v(x)$. Notice that in the limit $x \to \infty$, the solution
(\ref{eq:psolution}) for the pressure approaches the form $p \to C_2
\alpha^\Gamma$ (obtained by integrating the equations of motion). 
In the opposite limit $x \to 0$, $p \to C_3 \alpha^\gamma$. In other
words, this solution for the pressure illustrates how the equation of
state smoothly transforms from static index $\Gamma$ (at early times
and/or large radii) to the dynamic index $\gamma$ (at late times
and/or small radii).

In this paper, we consider only positive values of time, since we take
the zero point of the time variable to be the onset of collapse. It is
mathematically possible to consider solutions for negative times, $t <
0$, which would correspond to negative values of $x$.  Solutions that
span the entire available range $-\infty < x < \infty$ are known as
``complete'' solutions (first obtained by Hunter 1977; see also
Whitworth \& Summers 1985). Note that the self-similar solutions of
this paper for the protostellar collapse phase ($t > 0$) cannot (in
general) be extended to the pre-stellar phase ($t < 0$), as they would
likely encounter a critical point and become singular. In other words,
dynamical evolution (using only collapse physics) in the $t < 0$
regime would not (in general) lead to the initial conditions used
here.  However, it is rather unlikely that molecular clouds will
evolve toward their centrally condensed initial configurations in a
self-similar manner subject to (only) the physics included in these
equations of motion.  Before the onset of collapse (for $t < 0$),
molecular clouds may evolve through the processes of ambipolar
diffusion (at least for small mass scales), shocks, turbulent
dissipation, cooling flows, condensation instabilities, and
cloud-cloud collisions. In addition, the cloud will most likely
initiate collapse before a completely self-similar equilibrium state
has been attained; the collapse will only become self-similar
asymptotically in time (i.e., the self-similar collapse solutions of
this paper are intermediate asymptotic solutions to the realistic
problem of the collapse of a finite cloud with finite central
density).  In any case, we limit this discussion to solutions with $0
< x < \infty$, sometimes called ``semi-complete'' solutions.

\subsection{Numerical Treatment} 

As a complement to the self-similar formulation, we also perform
numerical simulations of the collapse for the case where $\Gamma =
\gamma = 1$. For the isothermal equations of state considered here, 
$P = a_s^2 \rho$, the force equation can be written as
\be
{\partial\over\partial t} (\rho u) + {1\over r^2}
{\partial\over\partial r} (r^2\rho u^2) = 
-a_s^2 {\partial\rho\over\partial r} -\rho {GM \over r^2} \;.
\ee 
This form of the force equation is solved numerically along with
equation (\ref{eq:fullcont}) using the second-order-accurate scheme
outlined in Boss \& Myhill (1992). This scheme invokes a variation on
the standard predictor-corrector technique, and allows for nonuniform 
grid spacing.  Our grid is defined by 160 logrithmically spaced radii
($r_i$) spanning four orders of magnitude from the inner to the outer
boundaries.  The cell interfaces ($r_{i\pm 1/2}$) are then set at the
midway points of these radii. The mass terms at each radius ($M_i$)
are advanced in time (denoted by the superscript $n$) using the scheme  
\be
M_i^{n+1} = M_i^n - F_i^n 4 \pi r_i^2 \Delta t \; ,
\ee
where the flux terms are defined as
\be 
F_i^n = \rho_i^n v_i^n + {\Delta t \over 2} \left[
\rho_i^n \left( {\partial u \over \partial t} \right)_i^n + u_i^n 
\left( {\partial \rho \over \partial t} \right)_i^n \right] \; ,
\ee
and the right hand terms in the above equations are
defined in Boss \& Myhill (1992).   

\section{GENERAL ANALYTIC RESULTS}  

Given the formulation of the collapse problem, we now find several
analytic results. These results are general in that they apply to the
entire class of solutions considered in this paper.

\subsection{Relationship Between the Similarity Transformation \\ 
and the Static Equation of State}  

First, we show that the initial density profile of the core determines
the type of similarity transformation which describes the subsequent
collapse of the core. To start, we assume that the initial equilibrium
configuration of the system can be described by a ``static'' equation
of state of the form 
\be
P = \kappa \rho^\Gamma \, . 
\ee
Since the similarity transformation determines the manner in which
both the pressure $P$ and the density $\rho$ must scale with time,
this new equation of state imposes an additional constraint on the
system.  In particular, the similarity transformation requires that
the scaling exponent $a$ (which has been left arbitrary thus far),
must have the value 
\be
a = \Gamma - 2 \, , 
\ee 
where the similarity variable is defined by $x = A t^a r$. 
We also obtain a required value for the corresponding 
numerical coefficient $A$, i.e.,  
\be
A = \kappa^{-1/2} (4 \pi G)^{(\Gamma - 1)/2} \, .
\label{eq:bigaspec} 
\ee
With this choice for $A$, the reduced static equation of state has the
form $p = \alpha^\Gamma$.  As mentioned before, however, this
coefficient $A$ can be rescaled by a dimensionless constant without
changing the physics (as long as that constant is propagated throughout 
the set of equations of motion).

Notice that for the special case in which the static and dynamic
equations of state are the same (i.e., $\Gamma = \gamma$), the index
$q = 0$ (see equation [\ref{eq:qdef}]). In this case, ${\cal C}_0 = 1
= {\cal C}_1$ and the equation of motion which defines the pressure
has the obvious solution $p = \alpha^\Gamma$. In the general case, the
index $q$ is given by 
\be
q = {2 (\Gamma - \gamma) \over 3 \Gamma - 4 } \, . 
\label{eq:qdef2} 
\ee 

\subsection{Hydrostatic Equilibrium and Overdensity}  

One useful reference state is that in which the core is in hydrostatic
equilibrium at $t=0$ when the collapse begins.  In the limit of large
values of $x$, we thus expect the core to have no velocity so that
$v=0$.  In this limit, $x \gg v$, the equations of motion have
solutions of the form 
\be
m(x) = {2 - \Gamma \over 4 - 3 \Gamma} 
\Biggl[ {2 \Gamma (4 - 3 \Gamma) \over 
(2 - \Gamma)^2} \Biggr]^{1/(2-\Gamma)} \, 
x^{(4 - 3 \Gamma) / (2 - \Gamma)} \, , 
\ee
\be
\alpha(x) = \Biggl[ {2 \Gamma (4 - 3 \Gamma) \over 
(2 - \Gamma)^2} \Biggr]^{1/(2-\Gamma)} \, 
x^{- 2/(2 - \Gamma)} \, . 
\ee  
The density profile of the hydrostatic equilibrium configuration is
thus a power-law with index $\muden \equiv - 2 / (2 - \Gamma )$. 
Notice that the static equation of state alone determines the initial
configuration (as expected).  Notice also that solutions of this form
make sense only for static equations of state with $\Gamma < 4/3$.
This latter result is, of course, well known from stellar structure
theory (e.g., Chandrasekhar 1939).  

In this paper, we allow for the initial states to have densities (and
mass profiles) that are heavier than that appropriate for hydrostatic
equilibrium. For general (static) equations of state, we let the
parameter $\overdense$ denote the overdensity, so that the initial
mass and density profiles are given by the above equations with a
leading factor of $\overdense \ge 1$.  For an isothermal (static)
equation of state, however, we follow previous authors (e.g., Shu
1977) and write $\rho (r) = \abig a_s^2 / 2 \pi G r^2$ so that the
starting profiles have the known forms $\alpha (x) = \abig / x^2$ and
$m(x) = \abig x$. In this case, $\abig = 2$ corresponds to hydrostatic
equilibrium and $\abig > 2$ represents overdense initial states. Note
that $\overdense = \abig/2$ for isothermal initial configurations.

It is useful to consider the density and mass profiles in terms of
physical quantities. Including the constants that carry dimensions and
the overdensity parameter, we can write the profiles in the form 
\be
M(r) = 4 \pi \overdense {2 - \Gamma \over 4 - 3 \Gamma} \, 
\Biggl[ {\kappa \Gamma \over 2 \pi G} {4 - 3 \Gamma \over 
(2 - \Gamma)^2} \Biggr]^{1/(2-\Gamma)} \, \, 
r^{(4-3 \Gamma) /(2-\Gamma)} \, , 
\ee 
and 
\be
\rho(r) = \overdense \Biggl[ {\kappa \Gamma \over 2 \pi G} 
{4 - 3 \Gamma \over (2 - \Gamma)^2} \Biggr]^{1/(2-\Gamma)} 
\, \, r^{-2/(2-\Gamma)} \, . 
\ee 

\subsection{Location of the Expansion Wave}  

We can derive a general expression for the location of the expansion
wave. If the initial state of the cloud core is in hydrostatic
equilibrium, then the boundary between the inner collapsing region and
the outer static region will be a critical point in the flow. The
mathematical manifestation of this critical point is that the matrix
of coefficients in the system of differential equations (\ref{eq:contmass} 
-- \ref{eq:contrho}) must vanish (at the critical point). This
condition can be written in the general form 
\be
\gamma = \alpha^{1 - \Gamma} (2 - \Gamma)^2 x_H^2 \, , 
\label{eq:exlocation} 
\ee
where we have used the fact that $v=0$ and $p = \alpha^\Gamma$ at the
head of the expansion wave $x_H$.  Using the solutions for the starting 
density profile derived above, we can eliminate the reduced density
$\alpha$ and find the following expression for the location $x_H$ of
the head of the expansion wave 
\be
x_H = \gamma^{(1 - \Gamma/2)} \, \, (2 - \Gamma)^{-1} \, 
\bigl[ 2 \Gamma (4 - 3 \Gamma) \bigr]^{-(1-\Gamma)/2} \, . 
\ee
Notice that this expression is valid for initial states that are in
hydrostatic equilibrium ($\overdense = 1$) and does not include the
terms appropriate for nonzero initial velocities. In our analytic
derivations, we use this expression as a useful benchmark for
estimating infall rates. For future reference, it is also useful to
have the (total) mass $m_H$ enclosed within the expansion wave. Using
the starting density profile and the expression for $x_H$, we find 
\be
m_H = 2 \Gamma \, \gamma^{(2 - 3 \Gamma/2)} \, 
\bigl[ 2 \Gamma (4 - 3 \Gamma) \bigr]^{-3(1-\Gamma)/2} 
(2 - \Gamma)^{-2} \, . 
\label{eq:mhdef} 
\ee

\subsection{Time Dependence of the Mass Infall Rate}  

The static equation of state also determines the time dependence of
the mass infall rate $\dot M$.  The infall rate determines the time
scales for the star formation processes.  In addition, since these
cores have no well defined mass scales, the mass infall rate is
the physical quantity of importance for star formation.  Using the
above results, we find that the mass infall rate has the form 
\be
{\dot M} = m_0 \, \kappa^{3/2} \, (4 - 3 \Gamma) \, 
(4 \pi G)^{3(1-\Gamma)/2} \, G^{-1} \, t^{3(1-\Gamma)} \, , 
\label{eq:mdotoft}
\ee 
where $m_0$ is a constant which represents the reduced mass $m(x)$ 
in the inner limit $x \to 0$. Since the infalling material always 
approaches ballistic conditions in this inner limit, the magnitude of
the reduced mass $m_0$ determines the infall rate. Notice that, except
for this parameter $m_0$, which is expected to be roughly of order
unity, the mass infall rate is completely specified by our scaling
transformation. The remainder of this paper is (mostly) devoted to 
the determination of $m_0$ for the cases of interest. 

For the case of an isothermal initial state, $\Gamma = 1$, and the
infall rate ${\dot M}$ is constant in time.  Notice that for equations
of state softer than the isothermal case $(\Gamma < 1)$, the mass
infall rate increases with time.  On the other hand, for harder
equations of state ($\Gamma > 1$), the mass infall rate decreases with
time. Since we expect most star forming regions to have initial core
configurations corresponding to soft equations of state, the general
trend is for mass infall rates to be increasing functions of time.

\subsection{Matching onto Inner Solutions: Effects of Rotation} 

In this section, we determine the most important effects of rotation
on the collapse solutions. For the cases of interest, the inner limit
of the spherical collapse flow approaches a ballistic (pressure-free)
form. However, the spherical approximation must break down in the
inner region of the flow where conservation of angular momentum plays
an important role and where a circumstellar disk forms. We thus
consider the inner limit of our spherical solutions as the outer limit
of the rotating non-spherical solutions which we calculate below.
This exercise in intermediate asymptotic analysis remains valid under
the following ordering of (radial) size scales:
\be
R_\ast \ll R_C \ll r_H \, . 
\ee
In this ordering constraint, the scale $R_\ast$ is the radius of the
forming star and defines the inner boundary of the collapse flow. The
scale $R_C$ is the centrifugal radius (defined more precisely below)
which roughly divides the spherical outer region of the flow from
the highly nonspherical inner region. Finally, the scale $r_H$ is the
head of the expansion wave and (again roughly) divides the outer core 
from the collapsing inner core. This latter division is clean only 
for hydrostatic initial states. 

Here, we present the form of the density profile resulting from the
collapse of slowly rotating cloud cores.  This calculation generalizes
the isothermal case studied previously (Terebey, Shu, \& Cassen 1984),
where the collapse solution in the outer portion of the core matches
smoothly onto an ``inner solution'' that can be described analytically
(Cassen \& Moosman 1981 and Ulrich 1976).  In this inner regime,
parcels of gas spiral inward on nearly ballistic trajectories with
nearly zero total energy (and conserved specific angular momentum).  
The resulting orbits are parabolic and are described by the equation 
\be
{\mu_0 (1-\mu_0^2) \over \mu_0 - \mu} = {1 \over \zeta} \, , 
\label{eq:mueq}
\ee
where $\mu_0$ is the cosine of the angle $\theta_0$ of the
asymptotically radial streamline (i.e., the parabolic fluid trajectory
that is currently passing through the position given by $\zeta$ and
$\mu \equiv \cos \theta$ initially made the angle $\theta_0$ with
respect to the rotation axis).  Although equation (\ref{eq:mueq}) is
cubic in $\mu_0$ and has three roots, only one solution has physical
significance.  The quantity $\zeta$ is defined by 
\be
\zeta \equiv {j_\infty^2 \over G M r} \, 
= {R_C \over r} \, , \ee
where $j_\infty$ is the specific angular momentum of parcels of gas 
currently arriving at the origin along the equatorial plane.   
We have followed previous authors in assuming an initial state 
that is rotating at a constant rotation rate $\Omega$, so that 
the quantity $j_\infty$ is given by 
\be
j_\infty = \Omega r_\infty^2 \, , \ee
where $r_\infty$ is the starting radius of the material that is
arriving at the origin at a given time.  This radius can be determined
by inverting the initial mass distribution $M(r)$, i.e., 
\be
r_\infty (M) = \Biggl[ {M \over 4 \pi \overdense} 
\Biggr]^{(2 - \Gamma)/(4 - 3 \Gamma)} \, 
\Biggl[ {2 \pi G \over \kappa \Gamma} \, \, (2- \Gamma)^\Gamma \, \, 
(4 - 3 \Gamma)^{1- \Gamma} \Biggr]^{1/(4 - 3 \Gamma)} \, \, . 
\ee

Putting all of these results together, we can find the centrifugal
radius for any given static equation of state. In general form, the
centrifugal radius can be written 
\be 
R_C = \Omega^2 \Bigl[ \overdense^{-4(2-\Gamma)} M^{4-\Gamma} 
G^{3(\Gamma-1)} (2 \kappa \Gamma)^{-1} (4-3\Gamma)^{7-4\Gamma} 
(2-\Gamma)^{4\Gamma-6} (4\pi)^{4\Gamma - 7} \Bigr]^{1/(4-3\Gamma)} \, ,  
\label{eq:rcgeneral} 
\ee
where the parameter $\overdense \ge 1$ is the factor by which the
initial states are denser than that required for hydrostatic equilibrium.
The result for three specific cases are given below. For isothermal 
equations of state, $\Gamma$ = 1, we obtain the familiar result  
\be
R_C = {\Omega^2 G^3 M^3 \over 16 a_s^8 } \, \sim t^3 \, , 
\ee
in the logatropic limit 
\be
R_C = {\Omega^2 M \over 2 \pi P_0 } \, \sim t^4 \, , 
\ee
where $P_0$ is the coefficient in the equation of state (which has
units of pressure in this limit). 

Using equation (\ref{eq:mueq}) in conjunction with conservation of
angular momentum and conservation of energy one 
can determine the velocity field (Cassen \& Moosman 1981). 
The density distribution of the infalling material can be obtained 
by applying conservation of mass along a streamtube (Terebey, Shu, 
\& Cassen 1984; Chevalier 1983), i.e., 
\be
\rho(r,\theta) \, v_r \, r^2
\sin\theta \, d\theta \, d\phi = - { {\dot M} \over 4 \pi}
\sin\theta_0 \, d\theta_0 \, d\phi .  
\label{eq:rotden} 
\ee
Combining the above equations allows us to write the density
profile in the form 
\be
\rho(r,\theta) = { {\dot M} \over 4 \pi (GM)^{1/2} }
\, r^{-3/2} \, \Biggl( 1 + {\mu \over \mu_0} \Biggr)^{-1/2}
\Biggl[ 1 + {2 \zeta} P_2 (\mu_0) \Biggr]^{-1}  , \ee
where $P_2(\mu_0)$ = $(3\mu_0^2 - 1)/2$ is the Legendre polynomial of
order two.  Notice that this form for the density distribution is
general in that it applies to all of the collapse calculations 
considered in this paper.  The properties of the collapsing core
determine the form of the functions $\dot M$ and $\zeta = R_C/r$ which
appear in the general form (\ref{eq:rotden}).

\section{COLLAPSE SOLUTIONS FROM ISOTHERMAL INITIAL STATES}  

The solution for the collapse of a singular isothermal sphere has been
found previously (Shu 1977) and has been used in numerous applications
of star formation theory. Most applications use the ``expansion-wave
collapse solution'' which starts from a hydrostatic initial state and
has zero initial velocity. To account for the observed inward flows in
starless cores, one can generalize the solution to include overdense
initial states (as considered in Shu 1977) and/or nonzero starting
velocities. In this section, we include both of these complications
and show the relationship between them. We also generalize the
collapse solution to include the effects of a dynamic equation of
state which differs from isothermal, i.e., we set $\Gamma = 1$ but
allow $\gamma \ne 1$.  For these solutions, the parameter $\kappa$
that appears in the static equation of state has units of speed
squared and can be written in terms of the isothermal sound speed,
i.e., $\kappa$ = $a_s^2$.

\subsection{Self-similar Solutions} 

For the case of an isothermal initial state, with static $\Gamma = 1$, 
the parameter $q = 2(\gamma - 1)$ and the reduced equations of motion 
take the form  
\be
(v - x) {dv \over dx} + {1 \over \alpha} {dp \over dx} 
= - {m \over x^2}  \, , \ee  
\be
(v - x) {1 \over \alpha} {d \alpha \over dx} + 
{dv \over dx} = 2 (1 - v/x) \, , \ee
where the reduced pressure $p(x)$ and mass $m(x)$ are 
given by 
\be
p = {\cal C} \, \alpha^\gamma \, m^{2(\gamma - 1)} 
\qquad {\rm and} \qquad m = x^2 \alpha (x - v) \, \, . 
\ee  

In this paper, we explicitly consider the starting states to exhibit
two generalizations from exact hydrostatic equilibrium. The cores can
be overdense, i.e., with densities larger than can be pressure
supported so that $\abig > 2$.  We also consider the possibility of
nonzero initial flows as characterized by the initial speed $\vin$. As
a result, we take the starting states to have the form 
\be 
\alpha = {\abig \over x^2} \Bigl[ 1 - {\abig-2 \over 2x^2} 
+ {\cal O} (x^{-4}) \Bigr] \, , 
\label{eq:alasymp} 
\ee 
and 
\be 
v = - \vin - {\abig - 2 \over x} - {\vin \over x^2} - 
{ (\abig - 2) 
\bigl[ (5 - 2 \gamma)/3 - \abig/6 \bigr] - 2 \vin^2 \over x^3 } 
+ {\cal O} (x^{-4}) \, . 
\label{eq:vasymp} 
\ee 
The density profile has the same form as considered previously for
overdense states (Shu 1977), whereas the velocity field picks up
additional terms due to $\vin \ne 0$ and $\gamma \ne 1$.  The limit of
hydrostatic equilibrium corresponds to $\abig \to 2$ and $\vin \to 0$,
so that $v \to 0$ and $\alpha \to 2/x^2$. 

For isothermal starting states (static $\Gamma$ = 1), we are left with
a three parameter family of collapse solutions, as specified by the
density parameter $\abig$, the inward velocity scale $\vin$, and the
index $\gamma$ of the dynamic equation of state. For each choice
$(\abig, \vin, \gamma)$, we get a different similarity solution, and,
in particular, a value of the reduced mass at the origin $m_0$ (which
specifies the mass infall rate). The standard infall-collapse solution 
(Shu 1977) has $m_0$ = 0.975 and corresponds to a particular point in 
this parameter space, namely (2,0,1); that same paper also presents 
solutions for points $(\abig,0,1)$, although they are not widely used 
in subsequent work. 

The first results of this paper are summarized by Figures 1 and 2. In
Figure 1, we plot the reduced mass $m_0$ as a function of the
overdensity parameter $\abig$ for various values of the index $\gamma$
appearing in the dynamic equation of state. Figure 1 shows that the
dynamic equation of state has only a modest effect on the mass infall
rate, whereas variations in the overdensity $\abig$ are much more
important.  Similarly, Figure 2 shows the variation of the reduced
mass $m_0$ as the initial inward velocity $\vin$ increases. In this
case, the dynamic equation of state affects the $m_0$ values more than
in the case of overdense states, but variation of the inward speed
$\vin$ has greater influence.

\subsection{Numerical Collapse Solutions} 

The numerical treatment of the collapse problem can be used to verify
the semi-analytic results and to study alternate collapse scenarios
that are not fully self-similar. For this numerical exploration, we
specialize to the case $\Gamma = 1 = \gamma$. We consider collapse
solutions for various values of the overdensity parameter $\abig$,
where $\abig = 2$ is the value appropriate for hydrostatic equilibrium
(where we adopt the notation of Shu 1977 for isothermal starting
states).  In addition, we allow for a non-zero infall velocity $\uin$
in the initial state.  As discussed previously, our self-similar
treatment considers nonzero velocities in the outer region with the
specific profile given by equation (\ref{eq:vasymp}) for isothermal
initial conditions. We note that current observations measure the
inward velocities at a limited range of local radii, so a variety of
different velocity fields (inward speed as a function of radius)
remain possible. Here, we numerically determine the infall rates for a
collection of isothermal spheres with nonzero outer velocities that
are constant (with radius). The speed $\uin$ and the overdensity
parameter $\abig$ are thus the two parameters that can be varied.

Figure 3 shows one representative model with $\abig$ = 2.0 (a density
profile in hydrostatic equilibrium) and initial inward speed $\uin =
a_s/2$ (a typical observed value). The top panel of Figure 3 shows
the density profile at the initial time (dotted curve) and four
subsequent times. Except for the overall normalization constant, this
solution is nearly identical to that found in the original
self-similar collapse solutions (compare with Figure 3 of Shu
1977). The second panel shows the velocity profile.  The inner regime
approaches the ballistic (free-fall) form, $|v| \sim r^{-1/2}$, as
expected. On the outside (large radii), the inward velocity attaches
smoothly onto the imposed boundary condition $v$ = {\sl constant}.
The numerically calculated infall rate 
quickly approaches the constant value predicted by the
similarity solution and stays fixed at that value.

Figure 4 illustrates that the flow is indeed self-similar (as
expected). Here, the collapse solutions for both the density (top
panel) and velocity field (bottom panel) at four different times are
rescaled according to the similarity transformation of \S 2 and then
plotted together.  The result is one smooth function for both density
and velocity, as expected for self-similar flow fields.

The resulting values of $m_0$ are shown in Figure 5 for a set of
models with varying overdensities (specified by $\abig$) and varying
infall speeds in the range $\vin = \uin/a_s$ = 0 -- 1.0.  Figure 5
shows that the mass infall rates (specified through $m_0$) increase
with increasing overdensity $\abig$ and with increasing initial inward
speed $\vin$. Furthermore, the functional dependence of $m_0$ on both
$\abig$ and $\vin$ is approximately linear so that increasing $\abig$
and increasing $\vin$ produce similar effects on the collapse
solutions.  A more detailed comparison of the effects of $\abig$ and
$\vin$ on the mass infall rate emerges from the analytical estimates
derived in the following subsection.
 
\subsection{Analytic Estimates} 

In this subsection, we derive analytic estimates for $m_0$ for the
case of initial configurations with static $\Gamma = 1$.  We consider,
separately, both the case of overdense initial states with $\abig > 2$
and initial states with nonzero inward velocities $\vin$ at the start.
To start, for simplicity, we set dynamic $\gamma = 1$. These estimates, 
by necessity, require approximations, but in the end they provide both
analytic understanding of the collapse physics and accurate analytic 
formulas for the integration constants $m_0$ and hence the infall rates. 

We first consider the case of overdense initial states. For a given
fluid layer at initial radius $r_0$, the inward velocity is zero at
$t=0$ (where $x \to \infty$), but an inward flow soon develops. Due to
the self-similar nature of the problem, the value of the radius $r_0$
does not matter. To estimate $m_0$, we assume here that the fluid
layer migrates slowly inward as a nonzero inward velocity develops. 
As the layer moves inward, the expansion wave moves outward. When the 
expansion wave passes the fluid layer on its way out, we assume that
the layer subsequently falls inward in a pressure-free manner.  For a
starting radius $r_0$, the fluid layer has a radial location at later
times given by 
\be
r^2 = r_0^2 - (\abig - 2) a_s^2 t^2 \, , 
\ee
where $a_s$ is the sound speed of the initial state. The expansion 
wave has radial location $r_H = a_s t$, so the expansion 
wave crosses the fluid layer at time $t_c$ given by 
\be
t_c = (r_0/a_s) (\abig - 1)^{-1/2} \, . 
\ee 
At this time, the layer resides at radius $r_c$ given by 
\be
r_c = r_H = a_s t_c = r_0 (\abig - 1)^{-1/2} \, . 
\ee 
From this location at $r_c$, the layer collapses in a pressure-free 
manner. The layer is already moving inward with speed 
\be
\uinss = {a_s^2 (\abig - 2) \over r_H} t_c \, = \, a_s (\abig - 2) \, , 
\label{eq:ustart} 
\ee 
where we have kept only the leading order term in the expansion. 
With this starting speed, the time required for the fluid layer 
to fall to the origin is given by 
\be
t_f = r_c^{3/2} \big[ 2 G M (r_0) \bigr]^{-1/2} \, f(\eta) \, , 
\ee
where we have defined
\be
f(\eta) \equiv 
(1-\eta)^{-3/2} \Bigl\{ \sin^{-1} (1-\eta)^{1/2} - 
(\eta - \eta^2)^{1/2} \Bigr\} \, , 
\label{eq:fdef} 
\ee
where the parameter $\eta \equiv \uinss^2 r_c / [2GM(r_0)]$ measures
the size of the initial velocity. The effect of $\uinss$ is thus
determined by the dimensionless function $f(\eta)$.  Here, the enclosed
mass is $M(r_0)$, i.e., all of the mass within the starting radius
$r_0$, even though the layer now resides at the smaller radius
$r_c$. As result, the relevant value of $\eta$ is given by
\be 
\eta = { (\abig - 2)^2 \over (2 \abig) (\abig - 1)^{1/2} } \, . 
\ee 
In the limit of hydrostatic equilibrium, $\abig \to 2$ and $\eta \to
0$, so that we recover the classical result $f(\eta) \to \pi / 2$.  
Since the inward speed cannot be larger than the free-fall value, 
this formula is only valid for $\abig \le \abig_{max} \approx$ 9.35 
(although we are interested in more modest values of $\abig$). 
The infall time can be written  
\be
t_f = f(\eta) \, {r_0 / a_s \over (\abig - 1)^{3/4} 
(2\abig)^{1/2} } \, . 
\label{eq:timeff} 
\ee 
The total time for the fluid layer to fall is then given by the 
sum $t_T = t_c + t_f$, i.e., 
\be 
t_T = {r_0 / a_s \over (\abig - 1)^{1/2} } + 
{ f(\eta) r_0 / a_s \over (\abig - 1)^{3/4} (2\abig)^{1/2} } \, . 
\ee 
By the time $t_T$ that the fluid layer reaches the origin, the
expansion wave has traveled out to $r_{H} = a_s t_T$. After some
algebra, the ratio of reduced masses can be written in the form 
\be 
{m_0 \over m_H} = \Bigl\{ (\abig - 1)^{-1/2} + f(\eta) 
(2 \abig)^{-1/2} (\abig - 1)^{-3/4} \Bigr\}^{-1}  \, . 
\label{eq:mzeroa} 
\ee 
Since $m_H$ = $\abig$, this estimate for $m_0$ is now completely
specified. Notice that for sufficiently large value of $\abig$ 
(namely, $\abig \gta 2.85$), the ratio $m_0/m_H$ can exceed unity. 
This occurs because the inward velocities outside the nominal 
location of the expansion wave move material inward faster than 
the outward-moving expansion wave can enclose more material. 

Figure 6 shows a comparison of our self-similar formulation, the
analytic estimate derived above, and results from our numerical
treatment.  For this comparison, all initial states are taken to be
isothermal (so that static $\Gamma = 1$) and the collapse flow is
assumed to be isothermal (so that dynamic $\gamma$ = 1). The resulting
values of the dimensionless mass $m_0$, which specifies the mass
infall rate, are shown as a function of the overdensity parameter
$\abig$.  Notice that all of the curves are in good agreement.  The
self-similar results (solid curve) have been determined previously
(see Table I of Shu 1977), but such overdense solutions have been
largely ignored. The numerical results (dotted curve) demonstrate that
the self-similar formulation produces the correct infall rates for
cores that start with the same initial conditions (here, density
profiles that are sufficiently centrally condensed -- see also \S
4.4). In addition, the analytic estimates are in excellent agreement
with both of the more rigorously determined values. The analytic
treatment makes two approximations, which tend to cancel out: We first
assume that once the expansion wave overtakes a fluid layer, the
material immediately becomes ballistic, and hence this assumption acts
to overestimate the infall rate. We also assume that the initial speed
in given by equation (\ref{eq:ustart}), which keeps only the leading
term in a series expansion, and hence tends to underestimate the
infall rate.  As a result, the dependence of $m_0$ on the overdensity
($\abig$) is well described by equation (\ref{eq:mzeroa}). In spite
of its derivation, the result (eq. [\ref{eq:mzeroa}]) provides an 
accurate and user-friendly formula to specify infall rates.

We now present an analogous calculation for the case of initial inward
velocities ($\uin \ne 0$). To isolate the effects of the velocity, we
fix the density coefficient to be that appropriate for hydrostatic
equilibrium so that $\abig = 2$.  For a fluid layer at starting radius
$r_0$, the inward speed is nearly constant at $\uin$, and the radial
location as a function of time is given by 
\be
r (t) = r_0 - \uin t \, . 
\ee 
Using the standard expression for the expansion wave, which is only 
an approximation in the present context (see eq. [\ref{eq:exlocation}]), 
we can find the time $t_c$ at which the expansion wave crosses the 
fluid layer, 
\be
t_c = r_0 (\uin + a_s)^{-1} \, , 
\ee
and the corresponding location $r_c$ of the layer, 
\be
r_c = r_0 a_s (\uin + a_s)^{-1} \, .
\label{eq:rcvinf} 
\ee 
As before, we assume that the fluid layer collapses subsequently in 
a pressure-free manner. The additional time $t_f$ (beyond $t_c$)
required for the fluid layer to reach the center is still given by
equation (\ref{eq:timeff}), with the intermediate crossover radius
$r_c$ given by equation (\ref{eq:rcvinf}) and the parameter $\eta$
given by 
\be
\eta = {\vin^2 \over 4 (1 + \vin) } \, . 
\ee
Keep in mind that $\vin = \uin / a_s$. Thus far, we have assumed that 
the fluid layer moves inward at constant speed $\uin$ until it crosses 
the expansion wave and then falls inward with no pressure forces. 
In actuality, the layer will accelerate a small amount before crossing 
the expansion wave, but will be held back (again, a small amount) 
by pressure forces; these two corrections thus tend to cancel. However, 
we can account for this complication by introducing a dimensionless 
parameter $b$ through the ansatz $\vwig = b \vin$. After some algebra, 
the ratio of reduced masses takes the form 
\be 
{m_0 \over m_H} = \Bigl\{ \Bigl[ {1 \over 1 + \vwig} \Bigr] 
+ {1 \over 2} f[\eta(\vwig)] \Bigl[ {1 \over 1 + \vwig} \Bigr]^{3/2} 
\Bigr\}^{-1} \, .
\label{eq:mzerov} 
\ee
The mass contained within the expansion wave $m_H = 2$ for these
configurations, so the mass $m_0$ is completely specified.

Figure 7 shows how this analytic formula compares with the results of
the self-similar formulation and the numerical treatment of the
collapse problem. All of the initial states are taken to have $\abig$
= 2, as well as $\Gamma = 1 = \gamma$.  The resulting values of the
reduced mass $m_0$ are shown as a function of the starting inward
speed $\vin = \uin/a_s$. The solid curve shows the results of the
self-similar calculation and the dashed curve shows the results of the
numerical treatment. The dotted curve shows the analytic estimate
derived above, where we set $b=1.175$ to take into account the (small)
acceleration of the fluid layer, relative to the simple assumptions of
the derivation.  With this correction, the analytic estimate is in
good agreement with both the numerical and self-similar results. 

These analytic derivations allow us to make an approximate
transformation between the effects of increasing the initial density
($\abig > 2$) and increasing the starting inward velocity ($\vin >
0$). If we equate the derived expressions for $m_0$ for the two cases
(eqs. [\ref{eq:mzeroa}] and [\ref{eq:mzerov}]), we find that to a
reasonable approximation $\abig$ and $\vin$ are related through the
cubic equation 
\be 
\abig^2 (\abig - 1) \approx 4 (1 + \vin)^2 \, . 
\label{eq:acube} 
\ee 
For a typically observed value of $\vin = 0.5$, for example, the
corresponding value of $\abig$ that produces the same infall rate is
$\abig \approx 2.5$. For this case, $m_0 \approx 2.1$ and the infall
rate is larger than that of the infall-collapse solution (Shu 1977) by
115 percent (a factor of 2.15); this result agrees with a recent
calculation for the collapse of a magnetized singular isothermal
toroid (Allen et al. 2003b). Observations indicate that inward motions
span a range 0.04 -- 0.1 km/s (Lee et al. 1998), which correspond to
either starting speeds in the range $\vin = \uin/a_s = 0.1 - 0.5$ or
overdensities in the range $\abig = 2.1 - 2.5$. The largest value of
$\vin$ that one should consider is $\vin = 1$, which limits the
overdensity parameter to the range $\abig \le 2.9$ (using eq. 
[\ref{eq:acube}]). 

\subsection{Relation to Observations} 

Given the three parameter family of collapse solutions found thus far,
we need to specify the portion of the parameter space
$(\abig,\vin,\gamma)$ that corresponds to observed molecular cloud
cores. Since the dynamic equation of state (specified by $\gamma$)
does not affect the mass infall rates as much as the other variables
(Figures 1 -- 7), we focus this discussion on cases with $\gamma = 1 =
\Gamma$. This paper is motivated by the realization that some starless
cores are observed to have subsonic inward motions (Lee et al. 1999,
2001) at roughly half the sound speed, i.e., $\uin \approx a_s/2$. As
mentioned previously, these observed motions could arise in two
different ways: (1) The inward speed could be part of the initial
conditions; in this case, we model the collapse using a starting
velocity $\vin \ne 0$. (2) The observed speeds could arise because the
core has already begun to collapse on the outside; in this case, we
model the collapse using overdense initial states ($\abig > 2$), which
have zero velocity at $t=0$ but soon develop subsonic speeds in the
outer regions (beyond the expansion wave). We show here that both of
these possibilities can be made consistent with current observations, 
although the $\vin \ne 0$ case is favored. 

Both scenarios ($\abig > 2$ or $\vin > 0$) are constrained by the
observation that the cores in question exhibit inward motions but do
not contain detectable infrared sources. The theory is constrained
because the observed inward velocity must be large enough and the
predicted protostellar luminsoity must be small enough. For overdense
initial states, the predicted inward speed $\uin$ at a given radius
$r_{obs}$ of observation is given by 
\be
\uin = (\abig -2) a_s^2 t / r_{obs} \, , 
\ee
where $t$ is the time since the onset of collapse. The luminosity 
$L_{th}$ of the central source can be written in the form 
\be
L_{th} = {\cal F} {G M {\dot M} \over R_\ast} = 
{\cal F} m_0^2 {a_s^6 \over G R_\ast} t \, ,
\ee 
where $\cal F$ is the fraction of the total available luminosity $L_0$
(see eq. [\ref{eq:luminosity}]) that is realized in the central source.
To match the observations, the predicted inward speeds must be as
large as those seen, $\uin \ge u_{obs}$. At the same time, the
predicted luminosity must be less than the observed limit, $L_{th}
\le L_{lim}$ (where $L_{lim}$ = 0.1 -- 1 $L_\odot$). The first of
these constraints implies a lower limit on the time $t$ since collapse
began, while the second constraint implies an upper limit on $t$. In
order to satisfy both constraints, the following bound must be met 
\be 
{\cal F} {\uin \over a_s} \le 
L_{lim} {R_\ast G \over r_{obs} a_s^5} 
{\abig - 2 \over m_0^2} \approx 0.05 \, 
{\abig - 2 \over m_0^2} \, \lwig \, , 
\ee
where $\lwig \equiv L_{lim}/(1 L_\odot)$, and the numerical constant
on the right hand side is obtained by assuming $r_{obs}$ = 0.1 pc,
$a_s$ = 0.2 km/s, and $R_\ast = 2 \times 10^{11}$ cm (Stahler et
al. 1981). The function $F(\abig) \equiv (\abig - 2) / m_0^2$,
considered as a function of $\abig$, has a maximum value of about
0.112 at $\abig \approx 2.5$ (found by evaluation), and we obtain the
constraint 
\be
{\cal F} {\uin \over a_s} \le 0.006  \, \lwig \, . 
\ee 
The observed inward velocities lie in the range $u_{obs}/a_s = 0.25 -
0.5$, so this constraint requires the luminosity to be a small
fraction of the total available luminosity, namely ${\cal F} \lta
0.02$. This tight constraint greatly limits the available parameter
space for the case in which the observed inward velocities are part 
of the collapse ($\abig > 2$). 

For the scenario in which the inward velocities are part of the
initial conditions, the observation that $\vin = \uin/a_s = 0.25 -
0.5$ is built into the boundary conditions and does not further
constrain the solution.  However, the requirement that the core does
not display a detectable infrared source implies a time constraint, 
\be 
{\cal F} t \le L_{lim} {R_\ast G \over a_s^6 m_0^2} \approx 
1.3 \times 10^4 {\rm yr} \, (\lwig / m_0^2) \, , 
\ee 
where we have used the same numerical values as before. In order
to account for $u_{obs}/a_s$ = 0.25 -- 0.5 = $\vin$, the implied
values of $m_0$ lie in the range $m_0$ = 1.6 -- 2.2.  As a result, we
obtain a constraint of the form ${\cal F} t \lta 5100$ yr. In order to
allow for a respectably long viewing time, say $t \sim 5 \times 10^4$
yr (about half the expected collapse time), we would need ${\cal F}
\lta 0.10$. 

Although both scenarios can be made consistent with observations, the
case of overdense initial states is much more constrained.  This
overdense scenario -- in which the observed inward motions are part of
the collapse itself -- requires extremely inefficient power generation
in the central source, ${\cal F} \lta 0.02$. These solutions are also
constrained in that they cannot be extended to $t < 0$ to become
``complete'' solutions (again, see Hunter 1977). In contrast, the
$\vin \ne 0$ scenario -- in which the observed inward motions are part
of the initial conditions -- is less constrained, but still requires
${\cal F} \lta 0.10$ under reasonable assumptions.

\section{COLLAPSE SOLUTIONS FROM NON-ISOTHERMAL INITIAL STATES} 

In this section, we consider the collapse of molecular cloud cores
with initial density profiles that are shallower than $\rho \propto
r^{-2}$ (the profile of the singular isothermal sphere). As discussed
in previous papers (e.g., Lizano \& Shu 1989, McLaughlin \& Pudritz
1997), these shallower density profiles arise from hydrostatic
equilibrium models of gaseous spheres with equations of state that are
softer than isothermal. In the present context, such states correspond
to static $\Gamma < 1$. The self-similar collapse of such spheres have
been considered previously for the case of $\gamma = \Gamma$ (see
Cheng 1977, 1978, and the appendices to McLaughlin \& Pudritz 1997),
although the focus has been on hydrostatic initial conditions with
$\vin = 0$.  In this section, we generalize the self-similar collapse
problem to include nonisothermal states that are either overdense
($\overdense > 1$) or contain nonzero inward velocities ($\vin > 0$).
Notice also that the motivation for considering equations of state
that are softer than isothermal applies primarily to the static
equation of state, and not necessarily to the dynamic equation of
state. We thus generalize these collapse solutions to include the case
$\gamma \ne \Gamma$.

\subsection{Self-similar Collapse Solutions with a Single Equation of State} 

We first consider the case where $\gamma = \Gamma$ and generalize
previous work to include non-zero velocities. As discussed earlier,
the observed inward motions can arise in two conceptually different
ways: Such velocities can be ``part of the collapse'' and hence would
imply overdense initial states with $\overdense > 1$; the velocities
could also arise as ``part of the initial conditions'' and would imply
$\overdense = 1$ and $\vin > 0$. In either case, we can obtain 
self-similar solutions using standard methods. The equation of motion 
reduce to the form  
\be 
y {d v \over dx} + \Gamma \alpha^{\Gamma -2} {d \alpha \over dx} = 
{y \alpha \over 4 - 3 \Gamma} + (\Gamma - 1) v \, , 
\ee
\be 
{dv \over dx} + {y \over \alpha} {d \alpha \over dx} = 
2(1 - v/x) \, , 
\ee 
where we have defined $y \equiv (\Gamma - 2) x + v$. In order to 
consider overdense states and nonzero initial inward motions, we 
must specify the outer boundary conditions. Keep in mind that the 
limit $x \to \infty$ corresponds to either $r \to \infty$ or $t \to 0$ 
because of the nature of the similarity transformation. The starting 
states are thus a generalization of the results given in \S 4.1 and take 
the form 
\be 
\alpha = \overdense \beta_0 x^{-2/(2-\Gamma)} \Biggl\{ 
1 - \Delta_0 {3 \Gamma - 2 \over 2(2-\Gamma)} \, \, x^{-2/(2-\Gamma)} 
\Biggr\} \, , 
\ee
\be
v = - \vin - \Delta_0 x^{-\Gamma/(2-\Gamma)} \, , 
\label{eq:vboundary} 
\ee
where we have defined the dimensionless coefficients 
\be
\beta_0 = \Biggl[ {2 \Gamma (4 - 3 \Gamma) \over 
(2 - \Gamma)^2} \Biggr]^{1/(2-\Gamma)} \, 
\qquad {\rm and} \qquad 
\Delta_0 = \overdense \beta_0 \, {2 - \Gamma \over 4 - 3 \Gamma} 
\Bigl[ 1 - \overdense^{-(2-\Gamma)} \Bigr] \, . 
\ee 
Notice that inclusion of the constant term $\vin$ in the velocity
expansion breaks the self-similarity. Solutions with $\vin \ne 0$ can
be found, but, unlike the case of isothermal solutions discussed in
the previous section, the results depend on the choice of outer
boundary condition (see below).

We note that the equations of motion can contain critical points, 
which can be considered as generalized sonic points. The condition 
for the flow to pass through a critical point take the form 
\be 
\bigl[ (\Gamma - 2) x + v \bigr]^2 = \Gamma \alpha^{\Gamma - 1} \, . 
\label{eq:crit} 
\ee 
Although the fluid fields are continuous at the critical points
(defined by eq. [\ref{eq:crit}]), the derivatives (here, with respect
to the similarity variable $x$) of the fluid fields need not be. As a
result, care must be taken when numerically integrating through
critical points. One can expand the functions in the neighborhood of
the critical points and use the resulting analytic forms to integrate
through the point. This treatment has been discussed previously in the
literature (e.g., see Shu 1977 and McLaughlin \& Pudritz 1997) and
need not be belabored here. In the present application, only solutions
starting from hydrostatic initial conditions actually go through the
critical points. For starting states that are even slightly overdense,
or ones that start with small initial inward velocities, the solutions
avoid the critical points altogether.

The basic results of this section are shown in Figure 8 and 9. Keep in
mind that the infall rate, the quantity of interest in this collapse
calculation, is given by the analytic expression of equation
(\ref{eq:mdotoft}), with the reduced mass $m_0$ determined numerically
(as shown in the Figures). Figure 8 shows the reduced mass $m_0$ as a
function of the index $\Gamma = \gamma$ for varying values of the
overdensity parameter $\overdense$. The results are plotted as the
ratio $m_0/m_H$, where $m_H$ is the mass enclosed within the expansion
wave for the case of hydrostatic equilibrium, i.e., with $\overdense$
= 1. During collapse, the overdense states develop a rapid inward flow
well beyond the nominal location of the expansion wave (which is most
meaningful for the hydrostatic case). As a result, the mass infall
rates become rather large, and vary with overdensity more sensitively
than in the case of the isothermal collapse calculations of the
previous section. Specifically, as the density increases by a factor
of 2 ($\overdense = 1 \to 2$), $m_0$ increases by a factor of
$\sim$1840 in the logatropic limit $\Gamma \to 0$, compared to only a
factor of 5.7 in the isothermal limit $\Gamma \to 1$. Notice also that
because of the rapid speed at large $r$, the ratio $m_0/m_H$ can be
greater than unity.

Figure 9 shows the corresponding results for the reduced mass $m_0$
for varying values of the fixed inward speed $\vin$. As before, the
results are presented in terms of the ratio $m_0/m_H$, where $m_H$ is
the hydrostatic value (which is constant for all $\vin$).  These
solutions have a more complicated interpretation than their isothermal
counterparts of the previous section. For isothermal initial
conditions, the reduced speed $v$ is just the physical speed $u$
scaled by the (constant) isothermal sound speed.  For nonisothermal
conditions, the physical speed $u$ = $w v(x)$, where $v$ is the
(dimensionless) reduced velocity and $w \equiv \kappa^{1/2} (4 \pi G
t^2)^{(1-\Gamma)/2}$ is (esssentially) the effective sound speed at
the head of the expansion wave. Unlike the isothermal case, however,
the scaling speed $w$ is time dependent. As a result, the solutions
are not self-similar and depend on the boundary conditions, i.e., on
the starting value $x_0$ of the similarity variable where integration
of the equations of motion begins. Nonetheless, solutions can still be
found (see Figure 9), but the interpretation is not as clean as
before. The solutions shown in Figure 9 have been calculated by
applying the outer boundary condition (eq. [\ref{eq:vboundary}]) at
$x_0 = 100$. In contrast to the self-similar isothermal case of the
previous section, the resulting values of $m_0$ depend on the choice
of $x_0$. Nonetheless, the curves shown in Figure 9 illustrate the
expected property -- that collapse solutions with initial inward
motions lead to larger infall rates than those found in the
hydrostatic limit.

\subsection{Estimate for $m_0$ for Hydrostatic Initial Conditions}  

In this section, we derive an estimate for the reduced mass $m_0$ at
the origin for the case of hydrostatic initial states. This estimate
is found by calculating the value of $m_0$ that would arise for a
pressure-free collapse. Since the pressure forces are small, this
calculation should provide a reasonable estimate of the true value of
$m_0$. The pressure forces are not zero, however, and retard the flow
so that the true value of $m_0$ will be smaller than given by this
estimate. We correct for this difference using a simple analytical fit. 

To start, we consider a fluid layer at an initial radius $r_0$.
Because of the self-similar nature of the problem, the value of $r_0$
does not matter.  For the collapse solutions of interest, the fluid
layer will remain at rest until the expansion wave reaches the radius
$r_0$; afterward, the layer begins to fall.  As the layer falls
inward, the total mass enclosed within its current radial position
remains constant and is given by the hydrostatic profile (\S 3.2)
evaluated at $r_0$, i.e., $M(r_0)$. 

The time $t_f$ required for a fluid layer to fall from an initial 
radius $r_0$ is given by 
\be
t_f = {\pi \over 2} r_0^{3/2} \big( 2 G M \bigr)^{-1/2} \, , 
\label{eq:tfone} 
\ee
where we have assumed no pressure forces so that this time scale is a
lower limit. Notice that if the fluid layer is not assumed to start 
at rest, but rather has an initial inward velocity $v_0$, then the 
leading numerical coefficient will be different (not $\pi/2$).

The time $t_c$ for the expansion wave to reach the initial
radius $r_0$ is given by 
\be
t_c = \bigl[ A r_0 / x_H \bigr]^{1/(2 - \Gamma)} \, , 
\ee
where $A$ is the coefficient in the similarity transformation 
given by equation (\ref{eq:bigaspec}). 
When the fluid layer reaches the origin, the total time $t_T$ 
that has elapsed is given by the sum 
\be
t_T = t_c + t_f \, . 
\ee 

The total mass in the central object (at the origin) is the mass
enclosed within the radius $r_0$.  The total mass enclosed with the
expansion wave is given by the hydrostatic mass profile (see \S 3.2)
evaluated at the location of the expansion wave radius at time $t_T$,
i.e., at 
\be
r_H = x_H A^{-1} t_T^{2 - \Gamma} \, . 
\ee
Thus, the ratio of masses is given by 
\be
{m_0 \over m_H} = \Biggl[ {r_0 \over r_H} 
\Biggr]^{(4-3\Gamma)/(2-\Gamma)} = \Biggl[ { t_0 \over t_T} 
\Biggr]^{(4-3\Gamma)} \, . 
\ee 
In this case, since the initial states are hydrostatic, $r_0$ = $r_H$ 
and the ratio $m_0/m_H$ is always less than unity (compare with the 
overdense states of \S 4.3). 

We can use the similarity transformation and the hydrostatic initial
state to write the pressure-free collapse time $t_0$ as a fraction of
the total time $t_T$. We also introduce a correction factor $g > 1$,
which can be a function of the equation of state, and which accounts
for the fact that the true infall time is longer than the pressureless 
free fall time calculated above.  After some algebra, the mass ratio 
can be written in the form 
\be
{m_0 \over m_H} = \Biggl\{ 1 + g {\pi \over 4} 
\bigl( \gamma/\Gamma \bigr)^{1/2} \, (2 - \Gamma)^{-1/2} 
\Biggr\}^{-(4 - 3 \Gamma)} \, . 
\label{eq:mzerogam} 
\ee 
Since we already have a closed form expression for the enclosed mass
$m_H$ (equation [\ref{eq:mhdef}]), we thus have a completely analytic
estimate for $m_0$.  Also, since this estimate was derived assuming no
pressure forces, this expression (with $g=1$) represents an upper
limit for the reduced mass $m_0$.  As an example, we consider the
limit of the singular isothermal sphere (Shu 1977) where $\gamma =
\Gamma = 1$.  For this case, $m_H$ = 2 and we obtain the estimate
$m_0$ = $8/(4 + \pi)$ $\approx$ 1.12, which compares reasonably well
with the calculated value of $m_0 = 0.975$. Although the expression
(\ref{eq:mzerogam}) is only accurate at the 10 percent level, it
illustrates how the mass infall rate (which is proportional to $m_0$
-- equation [\ref{eq:mdotoft}]) varies with the equation of state
(mostly the static index $\Gamma$).

Figure 10 shows the values of the central mass $m_0$ (which sets the
mass infall rate) as a function of the index $\Gamma$ in the static
equation of state for the simplest case in which the dynamic $\gamma$
is the same as the static $\Gamma$. The solid curve shows the results
obtained from solving the self-similar equations of motion (this
result has been obtained previously -- see McLaughlin \& Pudritz
1997). The dashed curve shows our analytic approximation for $m_0$,
where we take into account the fact that pressure slows down the
infall so that the time required to fall from a given starting radius
is longer by a factor of $g$ (see equation [\ref{eq:mzerogam}]). This
effect is greater for softer equations of state, so we take $g =
g(\Gamma) = (4/3) (2 - \Gamma)^{9/10}$, where the index of 9/10 is
found by fitting the numerically determined curve. The result provides
almost an exact fit.

\subsection{Different Dynamic and Static Equations of State} 

Finally, we consider collapse flow with nonisothermal initial states
and varying choices for the dynamic equation of state.  Figure 11
shows the result of keeping a fixed static $\Gamma$ while allowing the
dynamic equation of state (dynamic $\gamma$) to vary. As shown, the
variation in the dynamic equation of state has relatively little
effect on the resulting reduced masses $m_0$, except when the dynamic
index $\gamma$ is much larger than the static index $\Gamma$. In this
latter regime, the pressure forces within the incoming flow become
large and the infall rates are suppressed.  This suppression is best
illustrated by considering a standard reference case where the static
index $\Gamma$ is arbitrary, but the dynamic equation of state is
chosen to be isothermal and fixed ($\gamma$ = 1). Figure 12 shows the
result for varying static indices $\Gamma$ and a fixed (isothermal)
dynamic equation of state. The ratio $m_0/m_H$ of the central mass
variable to the mass within the fiducial expansion wave (this ratio
determines the mass infall rate) is plotted as a function of the
static index $\Gamma$ (as before, the denominator uses the hydrostatic
value of $m_H$ with $\overdense$ = 1).  The solid curves show the case
of $\gamma = \Gamma$ for two choices of the overdensity ($\overdense$
= 1.5 and 2). The dashed curves show the resulting values of $m_0/m_H$
for the same starting conditions, and the same static $\Gamma$, but
with dynamic $\gamma$ = 1. This generalization to $\gamma \ne \Gamma$
leads to increased pressure forces during the collapse (relative to
the case of $\Gamma = \gamma$) and the resulting values of $m_0$ are
smaller.  The discrepancy grows with decreasing static index $\Gamma$.

Before leaving this section, we note that the mathematical formulation
of the collapse problem used in this paper (\S 2) does not allow 
self-similar solutions for the case with $\Gamma \to 0$ and $\gamma
\ne \Gamma$. In other words, in the logatropic limit, collapse with
other dynamic equations of state remains undefined. One manifestation
of this problem can be found in equation (\ref{eq:psolution}) for the
reduced pressure. In the limit $\Gamma \to 0$ ($\gamma \ne 0$), $p \to
constant$ instead of the required limiting form $p = \log \alpha$. The
class of solutions found in this paper are known mathematically as
self-similar solutions of the first kind (Barenblatt 2003). Although
the case of $\Gamma \to 0$, $\gamma \ne 0$ does not allow for a
self-similar solution of the first kind, it remains possible that the
problem will admit self-similar solutions of the second kind, also
known as incomplete similarity (Barenblatt 2003; Zeldovich 1956; von
Weizs{\"a}cker 1954). This complication is beyond the scope of this
paper and will be left for future consideration.

\section{CONCLUSION}  

\subsection{Summary of Results} 

In this paper, we have explored a wide variety of collapse solutions
for the collapse problem in star formation. These solutions describe
the collapse of idealized molecular cloud cores and/or the collapse of
kernels, the subcondensations within large cores that form stellar
groups and clusters.  Recent observations of star forming regions
(e.g., Lee et al. 1999, 2001) indicate that the outer boundary
condition for the collapse flow is not static; instead, the collapse
flow must match smoothly onto a subsonic, but finite, velocity field
at large radii.  This paper shows that a wide variety of self-similar
solutions have small but finite velocity fields at large radii, and 
we argue that solutions of this type provide a better description of
protostellar collapse. The most important ramification of this change
in the outer boundary condition is that the mass infall rates will be
larger for cores with nonzero flows.  We have quantified this change
using the semi-analytic technique of finding similarity solutions; we
have verified this result using a full numerical treatment of collapse
and have derived analytic bounds/estimates to provide greater physical
understanding of the process. All three methods are in good agreement
and show how the infall rates depends on the parameters of the
problem. Our specific results are listed below:

\noindent {[1]} In general, the equation of state that describes the
thermodynamics of the collapse flow -- that which determines the
evolution of entropy -- need not be the same as the effective equation
of state that sets up the initial configuration. For the case of
polytropic equations of state, this generalization means that the
dynamic index $\gamma$ need not be the same as the static index
$\Gamma$. This paper shows that similarity solutions can be found for
the more general problem containing two equations of state. In
addition, although previous work has focused on collapse flows that
match onto static outer boundary conditions, this paper shows that
gaseous spheres with inward flows at large radii also can be described
by similarity solutions; we parameterize this latter generalization by
introducing the reduced speed $\vin$ (the speed at $x \to \infty$). 
Including the overdensity parameter $\overdense$, which has been
considered previously, this paper shows that self-similar collapse
solutions exist for a four parameter family of conditions, where
points in this parameter space can be written in the form $(\Gamma,
\gamma, \overdense, \vin)$.

\noindent{[2]} The initial hydrostatic equilibrium configuration is
determined by the static equation of state. In order for equilibrium
configurations to exist, the index of the static equation of state
must satisfy the constraint $\Gamma < 4/3$. The power-law index
$\muden$ of the density profile is given by 
\be
\muden = - {2 \over 2 - \Gamma} \, . 
\ee
This result shows that the equilibrium density distribution becomes
less centrally concentrated as the static equation of state
becomes softer (as the static index $\Gamma$ decreases).

\noindent{[3]} The time dependence of the mass infall rate also 
depends only on the static equation of state and is given by 
\be{\dot M} \sim t^{3 (1 - \Gamma)} \, . \ee
As a result, static equations of state that are stiffer than
isothermal produce mass infall rates that decrease with time.
Similarly, static equations of state that are softer than isothermal
lead to mass infall rates that increase with time.  Since we expect
the equilibrium configurations of actual cloud cores to correspond to
the softer equations of state, the mass infall rate should either
remain constant (as for the isothermal case) or increase with time.
This result holds for any value of dynamic $\gamma$.

\noindent{[4]} Collapse solutions which approach a ballistic
free-fall form in the inner region ($x \ll 1$) exist for all dynamic
equations of state with indices $\gamma < 5/3$.  For these solutions,
the reduced fluid fields approach the asymptotic forms
\be
m \sim m_0 \, ,  \qquad v \sim - \sqrt{2 m_0} x^{-1/2} \, , 
\qquad \alpha \, \sim \, (4 - 3 \Gamma) \, \sqrt{m_0 / 2} \, 
x^{-3/2} \, , 
\ee 
in the limit $x \to 0$. The reduced mass $m_0$ sets the magnitude 
of the mass infall rate. As shown by equation (\ref{eq:mdotoft}), 
the infall rate is determined, except for the reduced mass, by 
the similarity transformation. 

\noindent{[5]} For collapse starting from an isothermal static
equation of state (which implies $\rho \sim r^{-2}$), we have solved
the full equations of motions (the original partial differential
equations) in addition to the similarity equations. These numerical
results provide important confirmations of the similarity approach:
(A) The full numerical treatment verifies the solutions
found semi-analytically; in particular, they predict the same mass
infall rates. (B) The numerical results show that non-singular initial
configurations, in particular those that have a central region of
constant density, approach the similarity solutions asymptotically in
time. Moreover, the convergence occurs relatively rapidly and smoothly, 
especially for the observationally preferred cases with $\overdense > 1$ 
and $\vin \ne 0$. 

\noindent{[6]} We have derived analytic estimates for the reduced mass
$m_0$ which sets the mass infall rate through equation
(\ref{eq:mdotoft}).  For isothermal starting states, we obtain
analytic expressions for $m_0$ as a function of overdensity $\abig$
(see equation [\ref{eq:mzeroa}]) and as a function of the initial
inward speed $\vin$ (see equation [\ref{eq:mzerov}]). These
expressions predict the same infall rates calculated from both the
numerical treatment (solving the partial differential equations) and
our semi-analytic approach (the similarity solutions) as shown in
Figures 6 and 7.  For the case of non-isothermal initial conditions,
we have derived an analytic expression for $m_0$ as a function of the
index $\Gamma$ (see equation [\ref{eq:mzerogam}]) for collapse with
$\gamma$ = $\Gamma$. The resulting expression is in good agreement
with the values calculated from the similarity solutions (Figure 10).

\noindent{[7]} For collapse solutions with $\gamma \ne \Gamma$, 
the infall rates are esssentially the same when the two indices are
comparable (see Figures 1 and 11). For soft equations of state 
$\Gamma \ll 1$, however, the infall rates are greatly suppressed for
relatively large values of dynamic $\gamma \sim 1$ (see Figure 12).
In the logatropic limit for the static equation of state, $\Gamma \to
0$, this formulation does not admit self-similar solutions of the
first kind for arbitrary dynamic $\gamma$.

\noindent{[8]} The observational motivation for this study is that
starless cores can exhibit inward motions at about half the sound
speed, $u_{in} \approx a_s/2$. These motions could arise from two
conceptually different sources. If the core is overdense $\overdense >
1$, then the entire core -- including the outer regions -- will begin
to collapse, with observable inward motions appearing on the outside
before an observable infrared source appears in the center.
Alternatively, the inward motions could be part of the initial
conditions -- they could be present at the effective zero point of
time (the start of collapse), presumably left over from the core
formation process. Both cases are compatible with current
observations, although the overdense scenario is more constrained. 
The net result, for both scenarios, is that the infall rate with
nonzero initial velocities is larger (than in the hydrostatic case) 
by a factor of two (see also Allen et al. 2003b). 

\subsection{Discussion} 

The specific results outlined in \S 6.1 represent a significant
mathematical generalization of the self-similar collapse problem,
which now includes the four parameter family of solutions given by
$(\Gamma, \gamma, \overdense, \vin)$. Moreover, the analytic
expressions derived for $m_0(\overdense)$, $m_0(\vin)$, and
$m_0(\Gamma)$ provide both a convenient means of obtaining the infall
rates (through equation [\ref{eq:mdotoft}]) and a path toward greater
physical understanding of the infall process.  This work has
additional astronomical implications, however, as we discuss next.

The generalized collapse solutions presented here are motivated by
astronomical observations which indicate that cores have inward
motions at large radii at the start of the collapse phase (Lee et
al. 1999, 2001). The first implication of these new solutions is that
molecular cloud cores will collapse with larger mass infall rates due
to these inward velocities. When matched to the observed boundary
conditions, our collapse solutions imply that the infall rates will be
a factor of two larger than previous estimates.

Even with the previous (smaller) infall rates, observed protostellar
objects have smaller luminosities than those predicted by theoretical
infall rates (e.g., Kenyon \& Hartmann 1995), unless a great deal of
the energy is stored as orbital motions. With the higher infall rates
now indicated, this luminosity problem is worse by a factor of two,
which means that even more energy must be stored as rotational energy
in the disk.  With increased infall rates, the predicted time scales
for star formation are shorter by this same factor of two.  As argued
from both observational (Myers \& Fuller 1992) and theoretical (Adams
\& Fatuzzo 1996) considerations, stars form on a time scale $t_{sf}
\approx 10^5$ years over a wide range of stellar masses. Although the
mean time is shorter, the prediction that the distribution of time
scales is much narrower than the distribution of stellar masses (the
IMF) remains the same.

The theoretical picture of star formation developed in the 1980s
considered molecular cloud cores to be (nearly) hydrostatic at the
beginning of the collapse phase. More recent observations of the
numbers of starless cores (e.g., Jijina et al. 1999) indicate that
core formation must take place more rapidly and is unlikely to involve
a purely quasi-static process. Some modification to the picture is thus
necessary. Many recent authors (see the review of MacLow \& Klessen
2004) have been exploring the opposite limit in which the star
formation process is fully dynamic and that cloud cores never really
form at all (at least not as separate, physically well-defined
entities).  The picture of star formation emerging from this study is
intermediate between these two extremes: Cores form relatively quickly
(through physical processes not calculated here), but still represent
the initial conditions for the following stage of protostellar collapse. 
With their rapid formation, the cores display non-zero inward motions
at large radii as observed. When this dynamic element is incorporated
into the collapse solutions (as calculated in this paper), the
subsequent collapse proceeds much as in previous models but with a
larger infall rate (due to the head start provided by the initial
velocities). As a result, the scenario retains the successes of the
standard picture (Shu et al. 1987), such as the spectral energy
distributions of young stellar objects (Adams et al. 1987) and their
corresponding observed emission maps (Walker et al. 1990). At the same
time, the picture can be made consistent with faster formation times
for cloud cores (Jijina et al. 1999) and the observed inward motions
at large radii (Lee et al. 1999).

This modification to the initial conditions -- inward velocities due
to either overdense starting states or nonzero $\vin$ as an initial
condition -- alleviates one theoretical difficulty associated with
protostellar collapse, but raises another.  On the positive side, the
inward velocities ensure that non-singular cores collapse smoothly,
and more rapidly approach the self-similar solutions. On the negative
side, given their dynamic formation, the cores no longer display a
clean separation between their formation stage and their collapse
stage.

In the (old) hydrostatic picture, molecular cloud cores live on the
edge between collapse and expansion. For example, in the isothermal
case (Shu 1977) where $\rho_0 = \abig a_s^2/4 \pi G r^2$ and $\abig =
2$ corresponds to equilibrium, cores with $\abig \ge 2$ will collapse
but cores with $\abig < 2$ will actually expand! If cores were formed
in states indistinguishable from the hydrostatic equilibrium
configuration with $\abig = 2$, then they would always be in danger of
expanding. In retrospect, it seems unlikely that the fate of a cloud
core -- to collapse or to expand -- should depend with hair-trigger
sensitivity on whether $\abig = 2 + \epsilon$ or $\abig = 2 - \epsilon$ 
(where $\epsilon \ll 1$). In contrast, cores with initial inward
velocities (as now observed) will collapse under a robust set of
circumstances.

The initial inward velocities also affect the issue of cores with
constant density centers (as observed by Ward-Thompson et al. 1999,
Bacmann et al. 2000, Tafalla et al. 2004). For flat-topped cores that
start close to hydrostatic equilibrium, realistic collapse flows
approach the self-similar form when the ratio $r_{out}/r_C \gta 20$
(Foster \& Chevalier; Whitworth et al. 1996). Roughly speaking, the
central (constant density) region collapses first, on a time scale
given by $\Delta \tau \approx (3 \pi/32 G \rho_0)^{1/2}$ (e.g.,
Spitzer 1978). For later times $t > \Delta \tau$, the flow approaches
the self-similar form as long as more of the cloud exists to subsequently
collapse (this condition holds for $r_{out}/r_C \gta 20$). For cores
with initial velocities, however, the time required for the central
(constant density) region to collapse is reduced (as calculated in
equation [\ref{eq:fdef}]) so that the requirement on the ratio
$r_{out}/r_C$ is less constraining. In order words, because observed
cores have initial inward velocities, the self-similar solutions
provide a better description of the collapse flow (than if the cores
were exactly hydrostatic).

An explicit demonstration of this trend is shown in Figure 13, which
plots the mass infall rates for a set of flat-topped cores as a
function of time. This set of numerical collapse calculations begins
with isothermal cores with sound speed $a_S$ = 0.2 km/s and an
extended central region of nearly constant density spanning $r_C
\approx 10^{16}$ cm.  The initial states are slightly overdense, with
$\abig$ = 2.2. The solid curve shows the resulting mass infall rate
(expressed in $M_\odot$ yr$^{-1}$) for a starting state with no
initial velocity. As found previously, the infall rate is essentially
zero to begin with, and then exhibits a sharp spike at the time when
the constant density region reaches the origin; the infall rate then
decreases with time and eventually approaches the asymptotic value
${\dot M} = m_0 a_S^3/G$ predicted by the similarity solution. The
dotted (dashed) curve shows the corresponding mass infall rates for
starting inward speeds of $\uin$ = 0.5 $a_S$ ($\uin$ = $a_S$). When
$\uin \ne 0$, the mass infall rates reach their peak values sooner and
they more rapidly approach their asymptotic values (which are within
about 10 percent of those predicted by the self-similar
calculations). In addition, for cores with $\uin \ne 0$, the maximum
infall rates are smaller and the asymptotic values are larger, so the
time variability is less severe.  More specifically, the ratio of the
peak infall rate to its asymptotic value is $\sim$8.4 for $\uin$ = 0
and only 2 -- 3 for $\uin \ne 0$. It has been suggested (e.g.,
Ward-Thompson et al. 1999) that the spike in the infall rate
corresponds to the Class O phase of protostellar evolution. Although
this work suggests that initial inward velocities smooth out the spike
to some degree, the infall rates are nonetheless larger early on and
can correspond to the Class 0 phase.

The new difficulty raised by the presence of initial inward speeds has
two components. First, the physical processes that explains core
formation must be able to account for the observed speeds. Second, the
core formation epoch must be more dynamic than considered previously,
and the distinction between the core formation stage of evolution and
the core collapse stage is not as clean.  While this paper clarifies
the collapse phase of star formation, the (earlier) core formation phase
requires further theoretical work.

Finally, one must keep in mind that this work generalizes spherical
collapse solutions in the absence of magnetic fields and angular
momentum. Nonrotating solutions are expected to apply at larger radii
where the flow is nearly spherical. The effects of rotation become
important in the inner part of the flow, where our solutions can be
matched onto inner solutions as outlined in \S 3.5. The inclusion of
magnetic fields into the collapse problem has been studied in a
variety of related contexts (see Galli \& Shu 1993ab; Li \& Shu 1996,
1997; Allen et al. 2003ab; Shu et al. 2004), including one collapse
calculation with nonzero initial velocities (see Figure 7 of Allen et
al. 2003b). With the inclusion of rotation (at rate $\Omega$) and
magnetic fields $\bf B$, one is left with a six parameter family of
initial conditions $(\Gamma, \gamma, \overdense, \vin, \Omega, {\bf B})$, 
where the velocity field ${\bf v}_\infty ({\bf r})$, the rotation
profile $\Omega ({\bf r})$, and the magnetic field strength ${\bf B}
({\bf r})$ can all be functions of the spatial coordinates. With this
paper (which explores collapse with $\vin \ne 0$ and $\gamma \ne
\Gamma$), the effects of these parameters on collapse have all been
considered individually. The challenge for the future is thus to
identify the portion of this parameter space that applies to
protostellar collapse and to properly incorporate all of the relevant
effects (simultaneously) into the collapse solutions.

\bigskip 
{} 
\bigskip 
\centerline{\bf Acknowledgments} 

We would like to thank Susana Lizano and Frank Shu for early
discussions concerning self-similar collapse and the effects of
varying equations of state. We also thank Daniele Galli and Joan
Najita for intermediate term discussions. Finally, we thank Gus
Evrard and Greg Laughlin for more recent input, especially regarding
numerical issues, and we thank the referee for a prompt and insightful
report.  This work was supported by the University of Michigan through
the Michigan Center for Theoretical Physics, by NASA through the
Astrophysics Theory Program, and by Xavier University through the
Hauck Foundation.

\newpage 
\centerline{\bf APPENDIX: EXISTENCE OF FREE-FALL SOLUTIONS} 

\medskip 
 
In this appendix, we show that ballistic free-fall solutions exist 
in the inner regime of the collapse flow for any dynamic equation 
of state with index $\gamma < 5/3$. 

We begin by taking the limiting form of the equations of 
motion for the case 
\be
x \ll |v|, \qquad x \ll 1 \, , 
\ee
which defines the inner regime of the collapse flow.
In this limit, the general expressions for the mass $m(x)$ 
and the pressure $p(x)$ reduce to 
\be
m(x) = {v x^2 \alpha \over 3a + 2 } \, , 
\ee
and 
\be
p(x) = {\cal C}_0 \alpha^\gamma (\alpha v x^2 / a)^q \, 
= \, {\cal C}_1 \alpha^\gamma \, m^q \, . 
\ee

Next, we note that the final equation of motion takes the form 
\be
{1 \over \alpha} {d \alpha \over dx}  + 
{1 \over v} {d v \over dx}  = - {2 \over x} \, , 
\ee
which can be immediately integrated to obtain 
\be
\alpha v x^2 = {\rm constant} \, . 
\ee
Comparing this result with the limiting form for the 
mass $m(x)$ shows that the enclosed mass always approaches 
a constant value in the inner limiting regime, i.e.,  
\be
m (x) \to m_0 = {\rm constant} \, . 
\ee
This result is quite general; it was obtained by assuming only 
that the coordinate $x$ is small (which defines what we mean by 
the inner regime of the flow) and by assuming that $x \ll |v|$.  
This latter assumption implies that we are considering only 
{\it infall} solutions in this argument.  As a general rule, 
{\it outflow} solutions to the equations of motion will also 
often exist.  The outflow solutions generally have $v \to 0$
and $x \to 0$ and hence the results of this current argument 
are not applicable. 

Given that the reduced mass approaches a constant value 
in the inner regime, the reduced pressure can be simplified 
also, i.e., 
\be
p (x) \to {\widetilde {\cal C}} \alpha^\gamma \, , 
\ee 
where we have defined a new constant ${\widetilde {\cal C}}$.
This result makes sense: it means that the pressure is 
given by a pure polytropic form in the inner limit, 
where the polytropic index is determined by the dynamic 
equation of state.  In this limit, the gas has lost all 
memory of the original ``static'' equation of state. 

The equation of motion (the force equation) reduces to 
the form  
\be
v {dv \over dx} + \gamma {\widetilde {\cal C}} 
\alpha^{\gamma - 2} {d \alpha \over dx} 
+ {m_0 \over x^2 } = 0 \, . 
\label{eq:apforce} 
\ee
Our objective is to show that a self-consistent 
free-fall solution exists for certain values of the 
index $\gamma$.  If a free-fall solution exists, then 
the velocity must approach the form 
\be
v = - (2 m_0)^{1/2} \, x^{-1/2} \, , 
\ee
where the minus sign indicates that the material is falling 
inward. Using this form for the velocity in the equation of
motion (\ref{eq:apforce}), we see that the first velocity term exactly 
balances the third gravity term.  This solution will exist 
provided that the second pressure term vanishes. 
Since $m \to m_0 \sim v x^2 \alpha$ in this limit, 
$\alpha \sim x^{-3/2}$.  The ratio ${\cal R}$ of the 
pressure term to the gravity term thus becomes 
\be
{\cal R} \sim  {\rm constant} \times x^{(5 - 3 \gamma)/2} \, . 
\ee  
This ratio vanishes in the limit of small $x$ provided that 
\be\gamma < 5/3 \, . \ee
Thus, self-consistent free-fall solutions exist in the 
inner limit for all dynamic equations of state with 
indices $\gamma < 5/3$. 

\vskip0.70truein
\centerline{\bf NOTE ADDED IN PROOF} 
\bigskip 

Related work by McKee \& Holliman (1999) shows that singular
isothermal spheres (with static $\Gamma$ = 1) that are initially in
hydrostatic equilibirum will not collapse if dynamic $\gamma > 4/3$.
On the other hand, this paper finds self-similar collapse solutions
for $\gamma < 5/3$ (e.g., see Figure 1).  The solutions presented here
use starting states that are overdense (not in hydrostatic
equilibrium) so they will always begin to collapse. In order for
continued collapse to take place, the gas must be able to get to the
inner, free-fall regime before pressure forces take over. In the usual
star formation scenario, the gas strikes a stellar surface, releases
energy, and collapse can continue. In the physical (not self-similar)
problem, inner boundary conditions thus help determine whether or not
sustained collapse takes place. As a result, for cases with dynamic
indices in the range $4/3 < \gamma < 5/3$, the self-similar collapse
solutions of this paper may not always provide good models for physical
collapse solutions that include additional scales (e.g., finite mass,
fixed outer boundary, finite central density). The determination of
which physical conditions allow continued collapse in this regime must
be left for a future research project. However, the majority of our
collapse solutions, those with dynamic $\gamma < 4/3$, are not
affected by this issue (we thank Chris McKee for useful discussions 
on this topic).

\newpage 
\centerline{\bf REFERENCES} 
\medskip 

\par\pp
Adams, F. C. 1990, ApJ, 363, 578

\par\pp
Adams, F. C., \& Fatuzzo, M. 1996, ApJ, 464, 256 

\par\pp
Adams, F. C., Lada, C. J., \& Shu, F. H. 1987, ApJ, 321, 788 

\par\pp
Allen, A., Li, Z.-Yi., \& Shu, F. H. 2003a, ApJ, 599, 363

\par\pp
Allen, A., Shu, F. H., \& Li, Z.-Yi. 2003b, ApJ, 599, 351 

\par\pp
Bacmann, A., Andr{\'e}, P., Puget, J.-L., Abergel, A., Bontemps, S., 
\& Ward-Thompson, D. 2000, A\&A, 361, 555 

\par\pp
Barenblatt, G. I. 2003, Scaling (Cambridge: Cambridge Univ. Press) 

\par\pp
Benson, P. J., \& Myers, P. C. 1987, ApJS, 71, 89   

\par\pp
Boss, A. P., \& Myhill, E. A. 1992, ApJSS, 83, 311

\par\pp
Cassen, P., \& Moosman, A. 1981, Icarus, 48, 353

\par\pp
Chandrasekhar, S. 1939, Stellar Structure (New York: Dover) 

\par\pp
Cheng, A. F. 1977, ApJ, 213, 537  

\par\pp
Cheng, A. F. 1978, ApJ, 221, 320  

\par\pp
Chevalier, R. A. 1983, ApJ, 268, 753

\par\pp
Ciolek, G. E., \& Basu, S. 2001, ApJ, 547, 272  

\par\pp
Curry, C., \& McKee, C. F. 2000, ApJ, 528, 734 

\par\pp
Curry, C., \& Stahler, S. W. 2001, ApJ, 555, 160 

\par\pp
Fatuzzo, M., \& Adams, F. C. 2002, ApJ, 570, 210   

\par\pp
Foster, P. N., \& Chevalier, R. A. 1993, ApJ, 416, 303 

\par\pp
Galli, D., \& Shu, F. H. 1993a, ApJ, 417, 220

\par\pp
Galli, D., \& Shu, F. H. 1993b, ApJ, 417, 243

\par\pp
Hunter, C. 1977, ApJ, 218, 834 

\par\pp
Jijina, J., \& Adams, F. C. 1996, ApJ, 462, 874 

\par\pp
Jijina, J., Myers, P. C., \& Adams, F. C. 1999, ApJ Suppl., 125, 161  

\par\pp 
Kenyon, S. J., \& Hartmann, L. W. 1995, ApJS, 101, 117 

\par\pp
Larson, R. B. 1969a, MNRAS, 145, 271

\par\pp
Larson, R. B. 1969b, MNRAS, 145, 297

\par\pp
Larson, R. B. 1985, MNRAS, 214, 379  

\par\pp 
Lee, C. W., Myers, P. C., \& Tafalla, M. 1999, ApJ, 526, 788 

\par\pp 
Lee, C. W., Myers, P. C., \& Tafalla, M. 2001, ApJ Suppl., 136, 703 

\par\pp 
Li, Y., Klessen, R. S.,\& MacLow, M.-M. 2003, ApJ, 592, 975  

\par\pp
Li, Z.-Y. 1999, ApJ, 526, 806 

\par\pp
Li, Z.-Y., \& Shu, F. H. 1996, ApJ, 472, 211L 

\par\pp
Li, Z.-Y., \& Shu, F. H. 1997, ApJ, 475, 237 

\par\pp
Lizano, S., \& Shu, F. H. 1989, ApJ, 342, 834

\par\pp
Mac Low, M.-M., \& Klessen, R. S. 2004, Rev. Mod. Phys., 76, 125 


\par\pp 
McKee, C. F., \& Holliman, J. H. 1999, ApJ, 522, 313 

\par\pp
McLaughlin, D. E., \& Pudritz, R. E. 1997, ApJ, 476, 750 

\par\pp
Motte, F., \& Andr{\'e}, P. 2001, A\&A, 365, 440 

\par\pp 
Mouschovias, T. Ch. 1976, ApJ, 207, 141

\par\pp  
Myers, P. C. 1998, ApJ, 496, L109 

\par\pp  
Myers, P. C. 2004, submitted to ApJ 

\par\pp
Myers, P. C., \& Benson, P. J. 1983, ApJ, 266, 309  

\par\pp
Myers, P. C., \& Fuller, G. A. 1992, ApJ, 396, 631 

\par\pp  
Myers, P. C., \& Lazarian, A. 1998, ApJ, 507, L157  

\par\pp
Nakano, T. 1984, Fund. Cosmic Phys., 9, 139

\par\pp
Penston, M. V. 1969a, MNRAS, 144, 425 

\par\pp
Penston, M. V. 1969b, MNRAS, 145, 457 
 
\par\pp
Shu, F. H. 1977, ApJ, 214, 488 

\par\pp
Shu, F. H. 1983, ApJ, 273, 202 


\par\pp
Shu, F. H., Adams, F. C., \& Lizano, S. 1987, A R A \& A, 
25, 23

\par\pp
Shu, F. H., Li, Z.-Y., \& Allen, A. 2004, ApJ, 601, 930  

\par\pp
Spaans, M., \& Silk, J. 2000, ApJ, 538, 115 

\par\pp
Spitzer, L. 1978, Physical Processes in the Interstellar Medium 
(New York: Wiley) 

\par\pp
Stahler, S. W., Shu, F. H., \& Taam R. E. 1980, ApJ, 241, 637  

\par\pp
Tafalla, M., Myers, P. C., Caselli, P., \& Walmsley, C. M. 2004, 
A\&A, 416, 191 

\par\pp
Terebey, S., Shu, F. H., \& Cassen, P. 1984, ApJ, 286, 529

\par\pp
Ulrich, R. K. 1976, ApJ, 210, 377

\par\pp
Vazquez-Semadeni, E., Canto, J., \& Lizano S. 1998, ApJ, 492, 596  

\par\pp
von Weizs{\"a}cker, C. F. 1954, Z. Naturforschung, 9A, 269 

\par\pp 
Walker, C. J., Adams, F. C., \& Lada, C. J. 1990, ApJ, 349, 515 

\par\pp
Ward-Thompson, D., Motte, F., \& Andr{\'e}, P. 1999, MNRAS, 305, 143 

\par\pp
Whitworth, A., \& Summers, D. 1985, MNRAS, 214, 1 

\par\pp
Whitworth, A. P., Bhattal, A. S., Francis, N., \& Watkins, S. J. 
1996, MNRAS, 283, 1061 

\par\pp
Wolfire, M. G., \& Cassinelli, J. 1986, ApJ, 310, 207

\par\pp
Wolfire, M. G., \& Cassinelli, J. 1987, ApJ, 319, 850

\par\pp 
Zeldovich, Ya. B. 1956, Sov. Phys. Acoustics, 2, 25

\par\pp  
Zweibel, E. G. 2002, ApJ, 567, 962

\newpage 
\begin{figure}
\figurenum{1}
{\centerline{\epsscale{0.90} \plotone{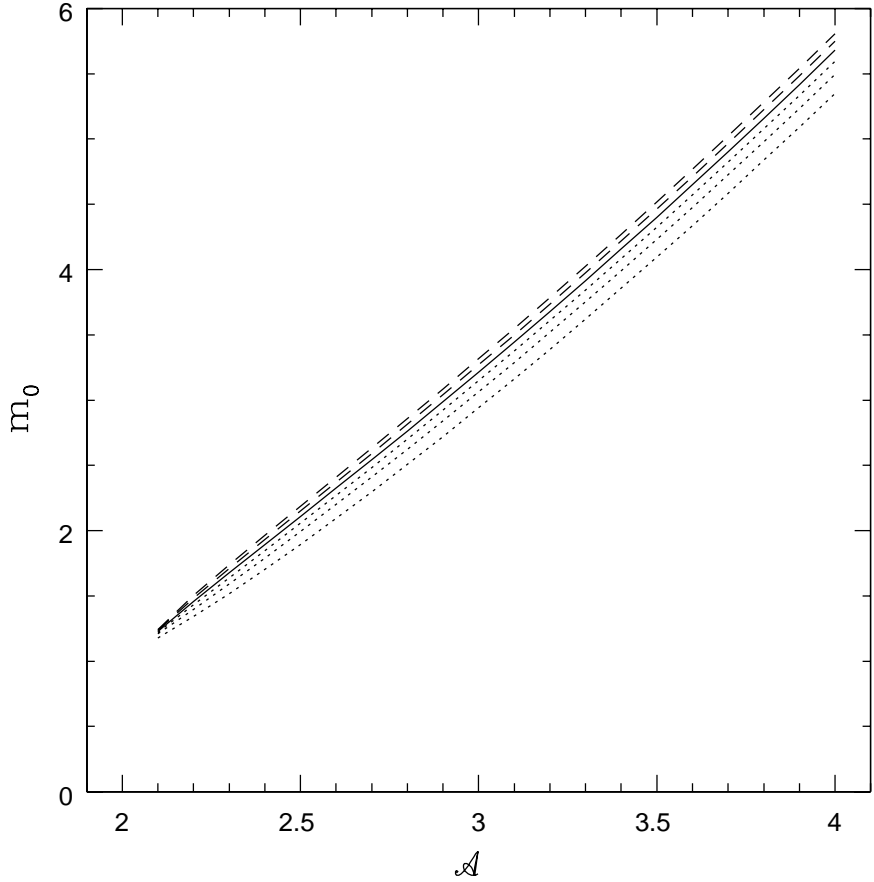} }}
\figcaption{For initial states with isothermal static $\Gamma$ = 1, 
the mass infall rate $\dot M$ = $m_0 a_s^3/G$. This plot shows the
variation of $m_0$ with the overdensity of the initial configuration
for various values of the dynamic $\gamma$. Hydrostatic equilibrium
corresponds to $\abig = 2$. The upper dashed curves show results for
$\gamma$ = 0.6 and 0.8 ($\gamma < 1$); the lower dotted curves show
results for $\gamma$ = 1.2, 1.4, and 1.6 ($\gamma > 1$); the solid
curve corresponds to $\gamma = 1$ (see Shu 1977). }  
\end{figure}

\newpage 
\begin{figure}
\figurenum{2}
{\centerline{\epsscale{0.90} \plotone{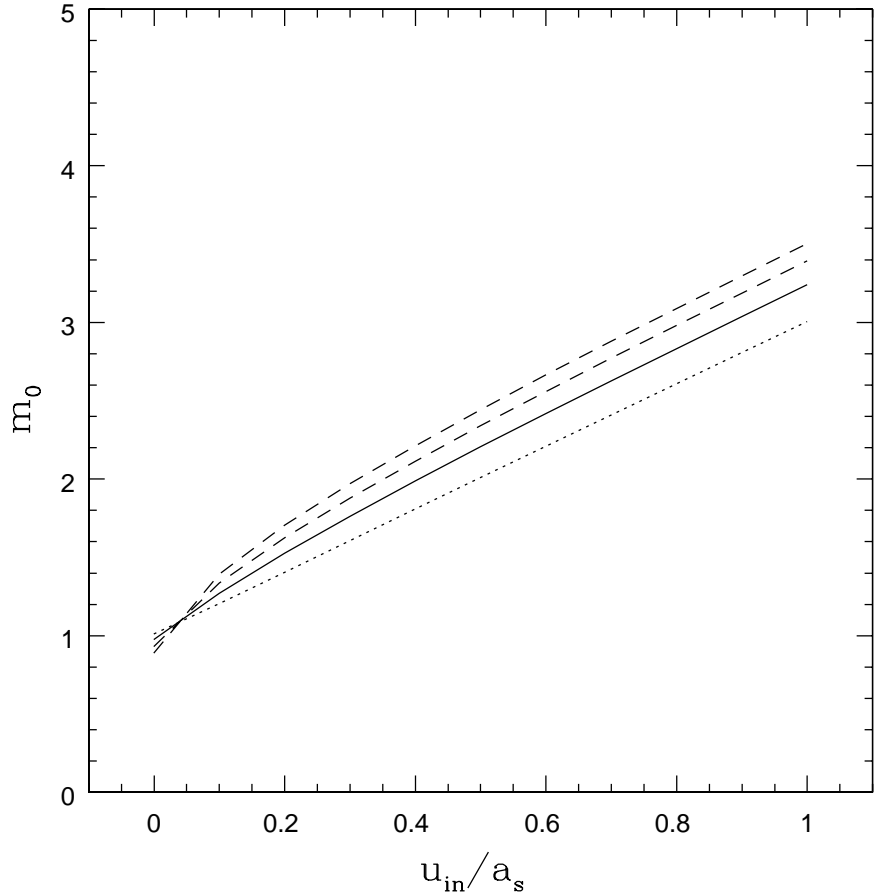} }}
\figcaption{For initial states with isothermal static $\Gamma$ = 1,
the mass infall rate $\dot M$ = $m_0 a_s^3/G$. This plot shows the
variation of $m_0$ with the initial inward velocity $\vin = \uin/a_s$
for various values of the dynamic $\gamma$. Hydrostatic equilibrium
corresponds to $\uin =0$. The upper dashed curves show results for
$\gamma$ = 0.6 and 0.8 ($\gamma < 1$); the lower dotted curve shows
results for $\gamma$ = 1.2; the solid curve corresponds to $\gamma = 1$. } 
\end{figure}

\newpage 
\begin{figure}
\figurenum{3}
{\centerline{\epsscale{0.90} \plotone{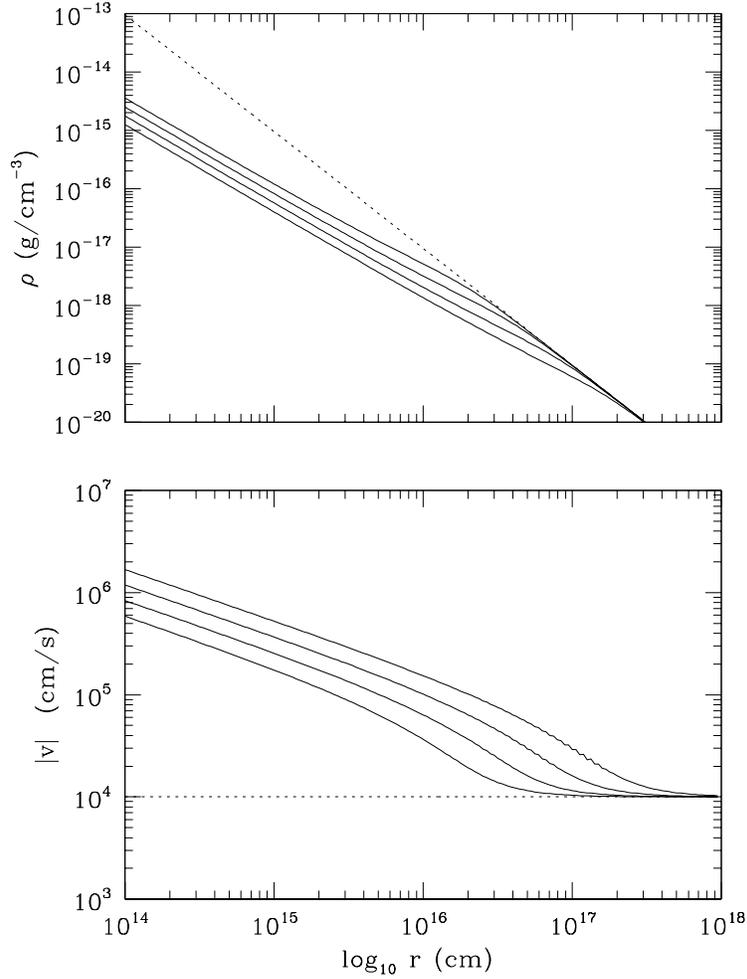} }}
\figcaption{This figure shows the numerically determined collapse
solution for a gaseous sphere with the density appropriate for
hydrostatic equilibrium ($\abig =2$) and an initial inward velocity of
magnitude $\uin = 0.5 a_s$. The top panel shows the resulting density
profile at four times (scaled to typical parameters for a molecular
cloud core). The bottom panel shows the corresponding inward velocity
profile at the same four epochs.  Notice that $v \to constant$ in the
limit $r \to \infty$ and the profiles have the same (self-similar)
form at smaller radii. }  
\end{figure}

\newpage 
\begin{figure}
\figurenum{4}
{\centerline{\epsscale{0.90} \plotone{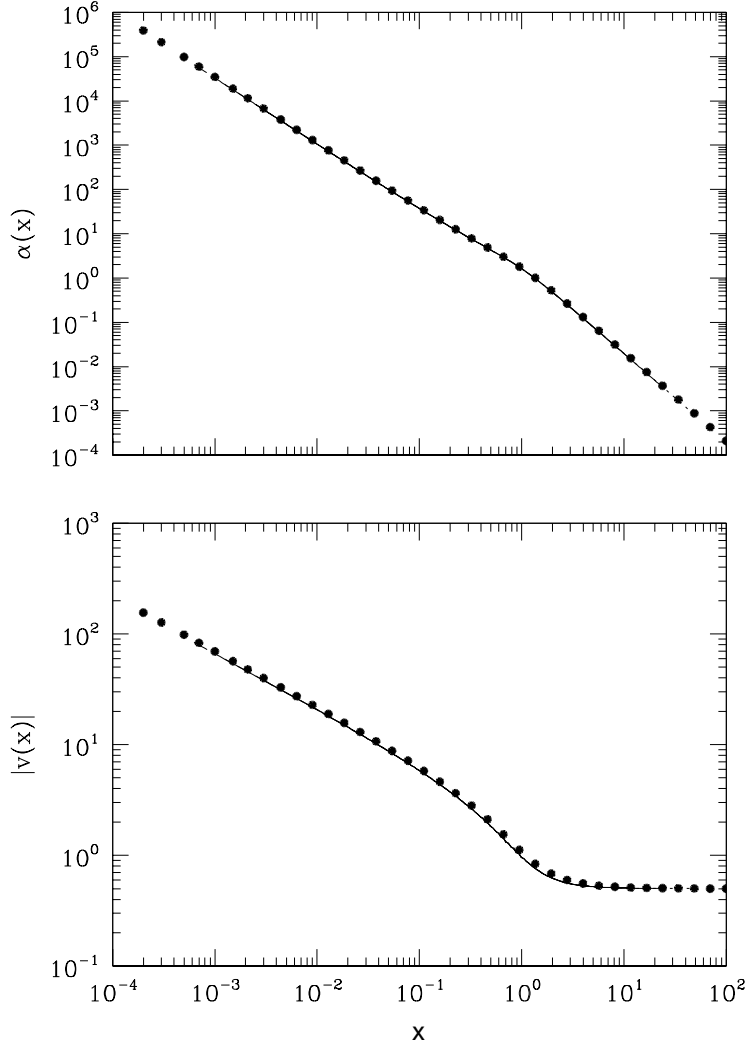} }}
\figcaption{This figure illustrates how the numerically determined
collapse solution for a gaseous sphere approaches a self-similar form.
The numerically determined solutions for both the density (top panel)
and velocity (bottom panel) are rescaled according to the (expected)
similarity transformation and then plotted on top of each other. The
result is one smooth function (for each panel) and hence the
numerically determined solutions at any given time are a stretched
version of the solution at an earlier time, in keeping with the
expectation of self-similarity. The corresponding self-similar 
solution, shown by the filled circles, is in good agreement. }  
\end{figure}

\newpage 
\begin{figure}
\figurenum{5}
{\centerline{\epsscale{0.90} \plotone{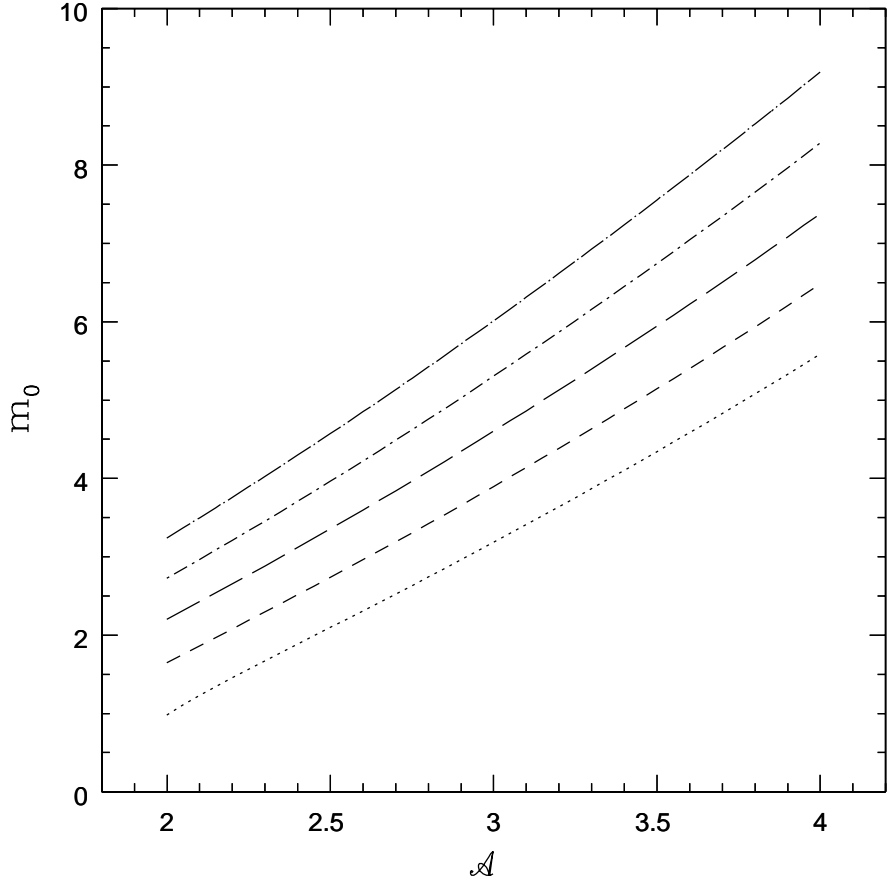} }}
\figcaption{Numerically determined infall rates shown as a function of
the overdensity parameter $\abig$ and the initial (inward) velocity.
All collapse models use isothermal conditions so that $\gamma$ = 1 = 
$\Gamma$. Each curve shows the numerically calculated values of $m_0$ 
as a function of $\abig$ for a given starting speed $\uin$. The 
lower curve shows the result for $\uin = 0$. The next four curves, 
in ascending order, show the results for $\uin/a_s$ = 0.25, 0.5, 
0.75, and 1.0, where $a_s$ is the isothermal sound speed. }
\end{figure} 

\newpage 
\begin{figure}
\figurenum{6}
{\centerline{\epsscale{0.90} \plotone{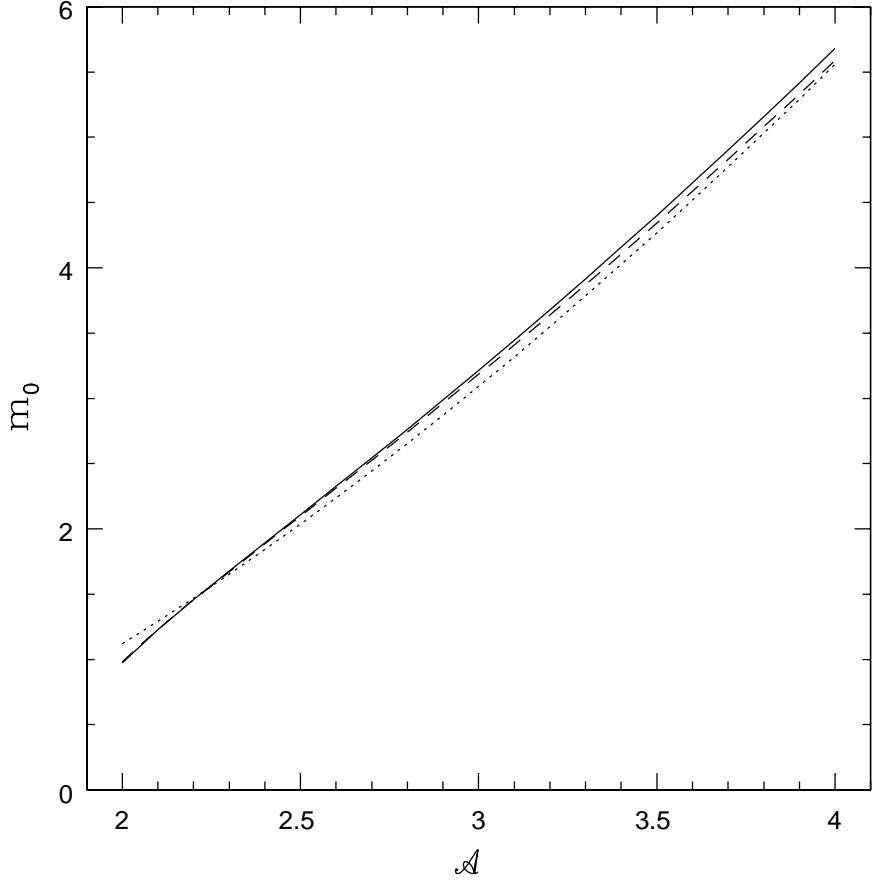} }}
\figcaption{Comparison of self-similar, analytic, and numerical
determinations of $m_0$, the dimensionless mass that specifies the
infall rate for isothermal initial conditions. The analytic estimate
for $m_0$ is shown by the dotted curve; the values calculated from the
self-similar equations of motion are shown by the solid curve; the
results from a numerical treatment (solving the partial differential
equations) are shown as the dashed curve. All models have initial
states with isothermal static $\Gamma$ = 1 and dynamic $\gamma$ = 1. 
Keep in mind that for all of these configurations, the mass infall
rate is given by $\dot M$ = $m_0 a_s^3/G$ so that $m_0$ specifies the
infall rate. }  
\end{figure}

\newpage 
\begin{figure}
\figurenum{7}
{\centerline{\epsscale{0.90} \plotone{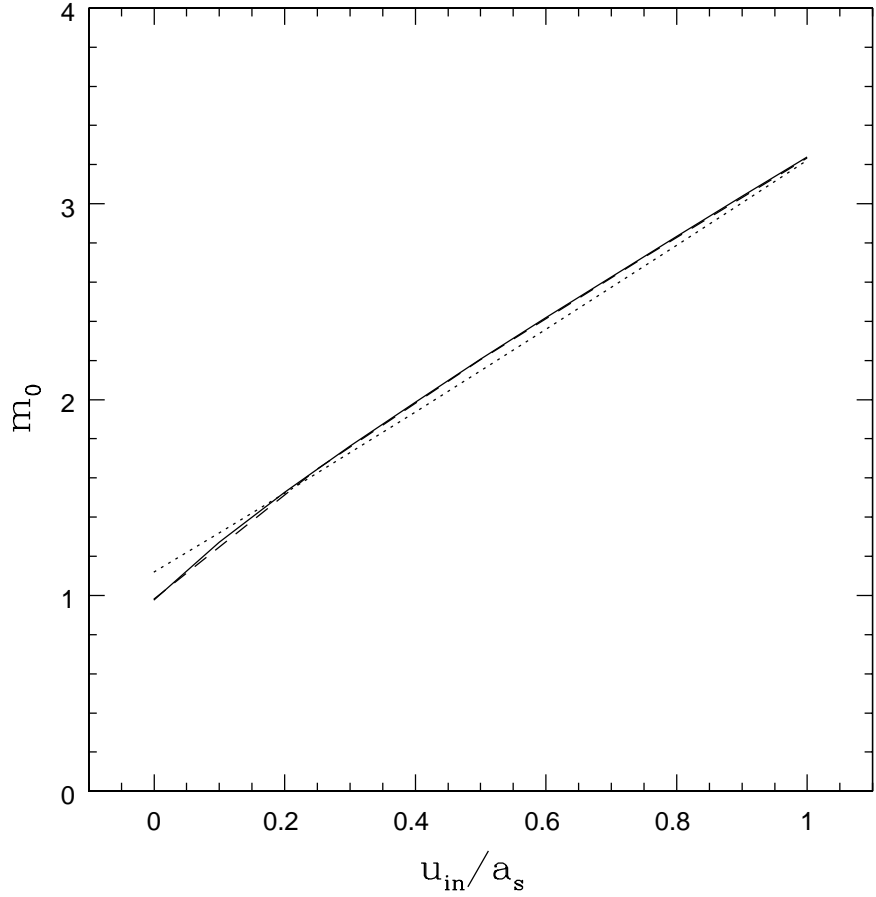} }}
\figcaption{Comparison of self-similar, analytic, and numerical
determinations of $m_0$ for varying initial inward velocities $\uin$
and for isothermal initial conditions.  The $m_0$ values calculated
from the self-similar equations of motion are shown by the solid
curve; the results from a numerical treatment (solving the partial
differential equations) are shown as the dashed curve.  The analytic
estimate for $m_0$ is shown by the dotted curve; to obtain this form,
we assume a fitting parameter that takes into account a slight (20
percent) acceleration of the flow before the expansion wave passes. 
All models use initial states with isothermal static $\Gamma$ = 1 
and dynamic $\gamma$ = 1. For all cases, the mass infall rate is 
given by $\dot M$ = $m_0 a_s^3/G$ so that $m_0$ specifies the infall 
rate. }  
\end{figure} 

\newpage 
\begin{figure}
\figurenum{8}
{\centerline{\epsscale{0.90} \plotone{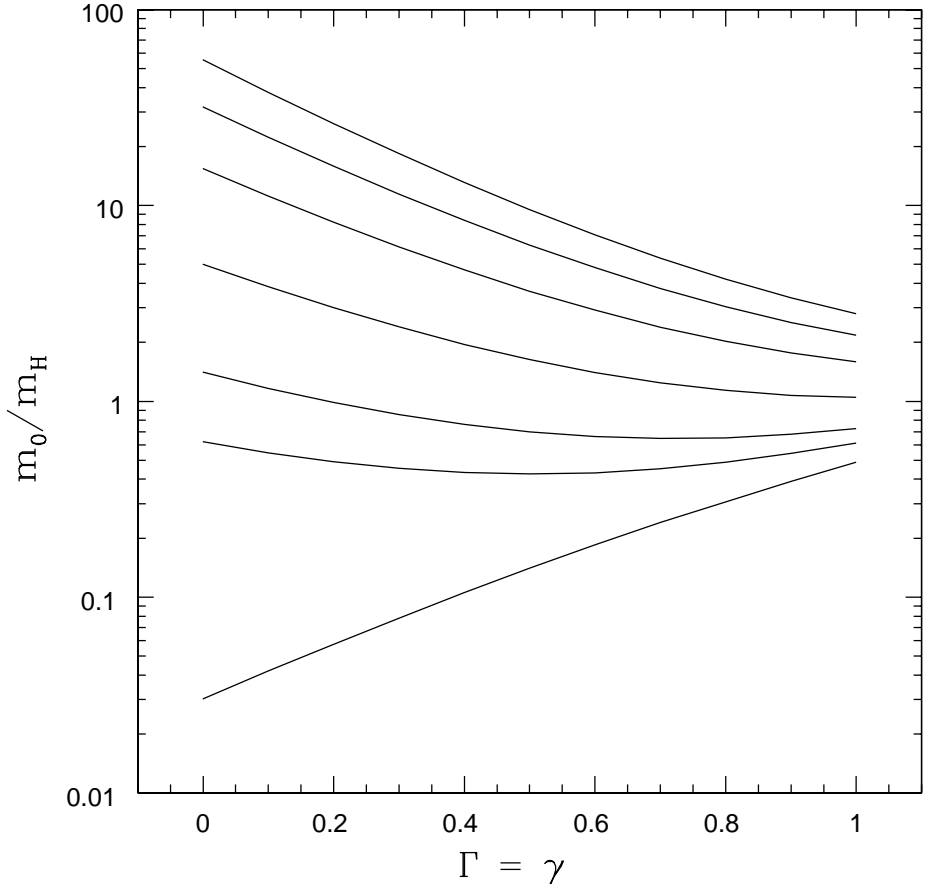} }}
\figcaption{Reduced masses $m_0$ as a function of equation of state
index $\Gamma = \gamma$ for varying values of the overdensity. The
results are expressed as the ratio $m_0/m_H$, where $m_H$ is the
reduced mass enclosed within the expansion wave for the case of
hydrostatic equilibrium (the reduced masses $m_0$ for each overdensity
is thus compared to the same $m_H$). The lowest curve shows the case 
of hydrostatic equilbrium with $\overdense = 1$. The subsequent curves
(proceeding upward) show the results for $\overdense$ = 1.05, 1.10,
1.25, 1.50, 1.75, and 2.0. Notice that the infall rate, which is
proportional to $m_0$, varies much more rapidly with the overdensity
$\overdense$ for smaller values of index $\Gamma$. }  
\end{figure} 

\newpage 
\begin{figure} 
\figurenum{9}
{\centerline{\epsscale{0.90} \plotone{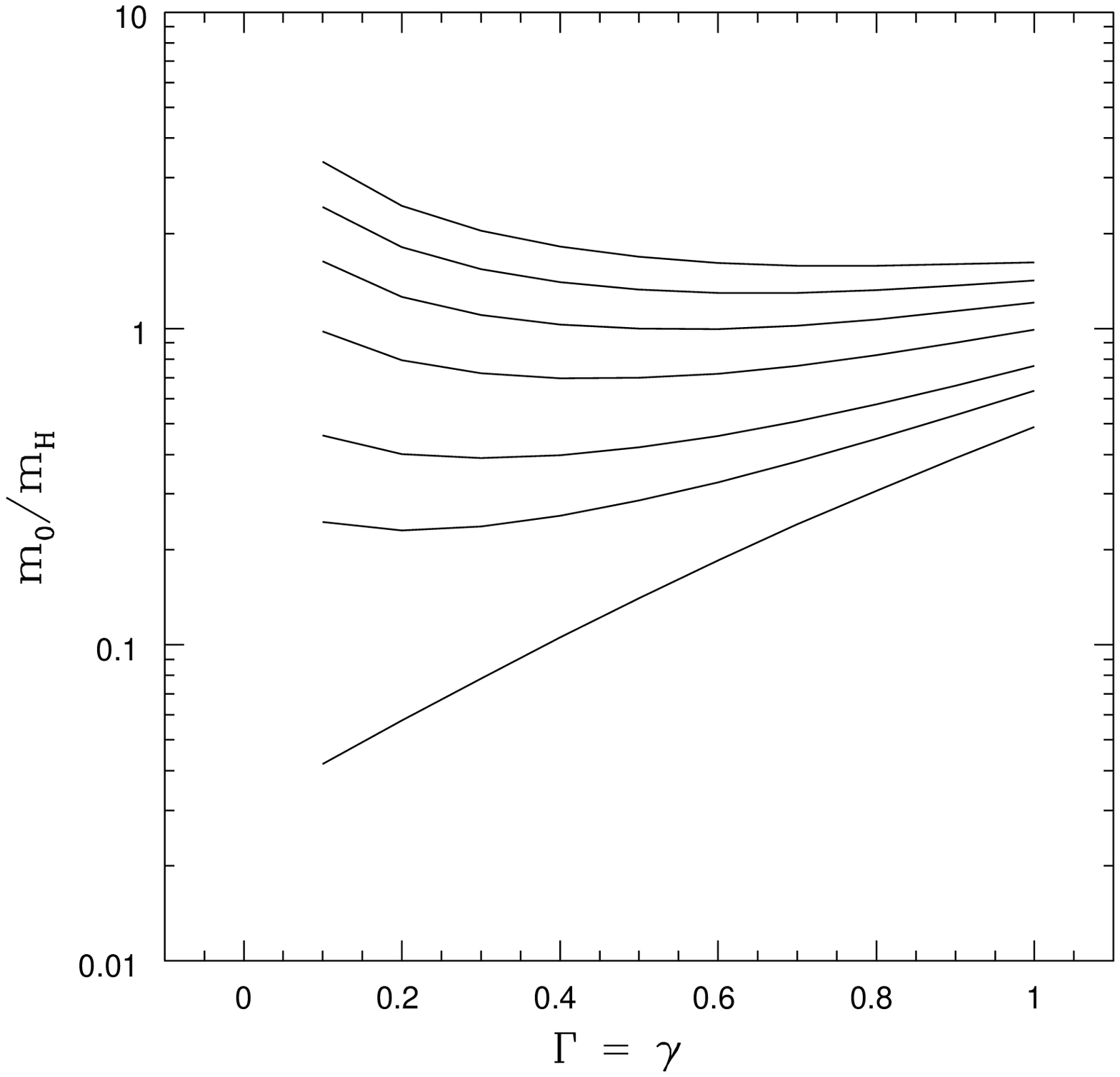} }}
\figcaption{Reduced masses $m_0$ as a function of equation of state
index $\Gamma = \gamma$ for varying values of the inward speed $\vin$
(where $v(x) \to \vin$ at large $x$). The results are expressed as the
ratio $m_0/m_H$, where $m_H$ is the reduced mass enclosed within the
expansion wave for the case of hydrostatic equilibrium (the reduced
masses $m_0$ for each case is compared to the same $m_H$). The lowest
curve shows the case of hydrostatic equilbrium with $\vin = 0$. The
subsequent curves (proceeding upward) show the results for $\vin$ =
0.1, 0.2, 0.4, 0.6, 0.8, and 1.0. Because the inclusion of $\vin \ne 0$ 
breaks the similarity, only the lowest curve is self-similar. The 
other curves of $m_0/m_H (\Gamma)$ depend on the value $x_0$ at which 
the outer boundary condition is applied; $x_0$ = 100 for the cases 
shown here. }  
\end{figure} 

\newpage 
\begin{figure}
\figurenum{10}
{\centerline{\epsscale{0.90} \plotone{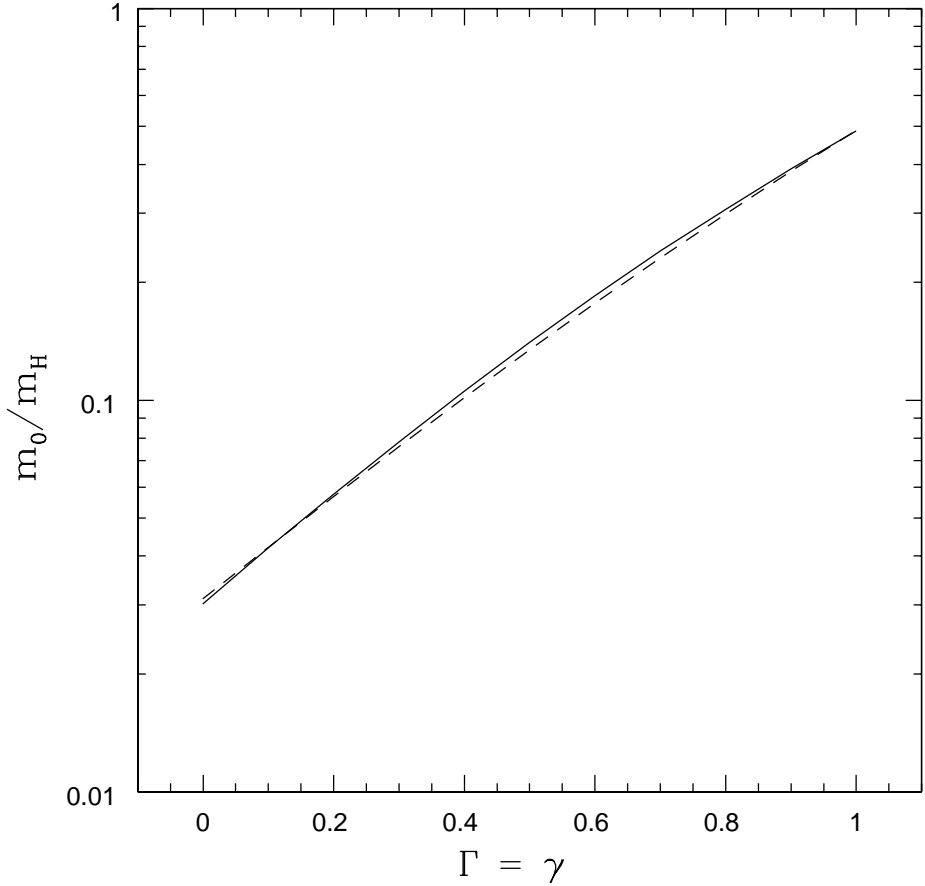} }}
\figcaption{Comparison of the analytic estimates for $m_0$ (dashed
curve) with the values calculated from the self-similar equations of
motion (solid curve) for initial states with varying values of the
polytropic index. In this case, the dynamic $\gamma$ is set equal to
the static $\Gamma$. The mass infall rate $\dot M$ is proportional to
$m_0$ (although the scalings depend on $\gamma = \Gamma$). The dashed
curve shows the analytic result of equation (\ref{eq:mzerogam}) with
$g(\Gamma) = (4/3) (2-\Gamma)^{9/10}$; this form takes into account
the result, found numerically, that polytropes with softer equations
of state approach free-fall conditions more slowly as $\Gamma$
decreases. } 
\end{figure}

\newpage 
\begin{figure}
\figurenum{11}
{\centerline{\epsscale{0.90} \plotone{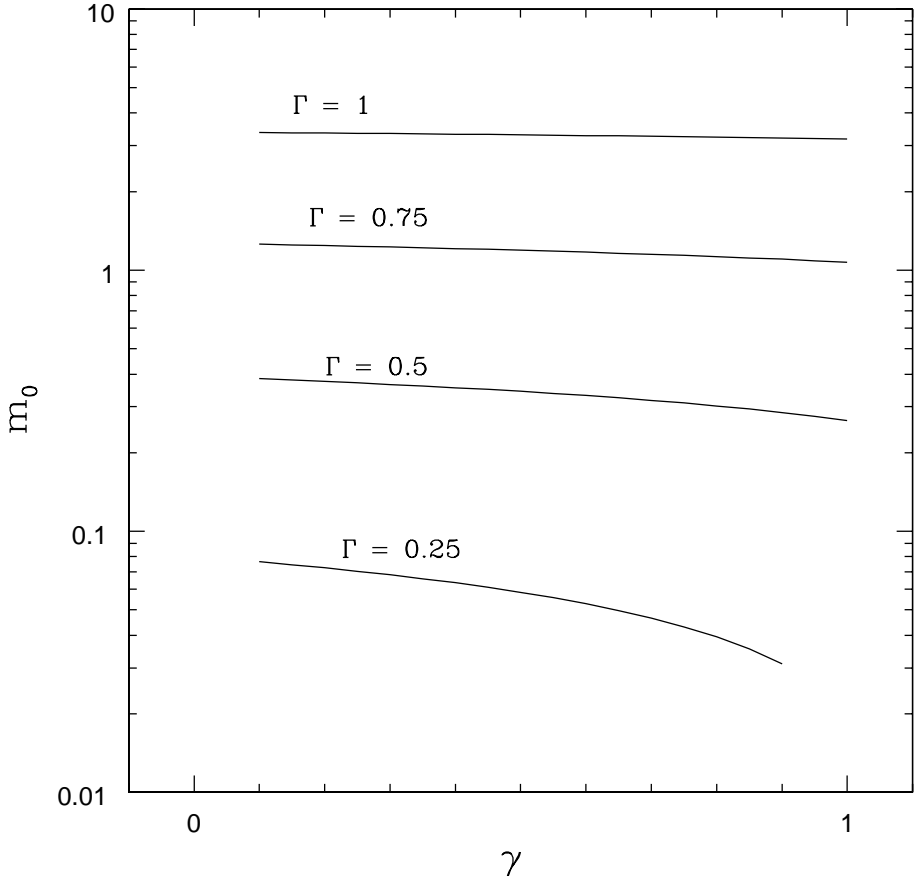} }}
\figcaption{The effect of dynamic $\gamma$ on infall rates. The 
various curves show the value of the reduced mass $m_0$ (which 
sets the infall rate) as a function of dynamic $\gamma$ for fixed 
values of the static equation of state, i.e., fixed values of 
static $\Gamma$ (as labeled). The variation in dynamic $\gamma$ 
has relatively little effect unless $\gamma \gg \Gamma$; in this 
latter regime, the reduced masses (and infall rates) become much 
smaller than in the case of $\gamma = \Gamma$. }  
\end{figure} 

\newpage 
\begin{figure}
\figurenum{12}
{\centerline{\epsscale{0.90} \plotone{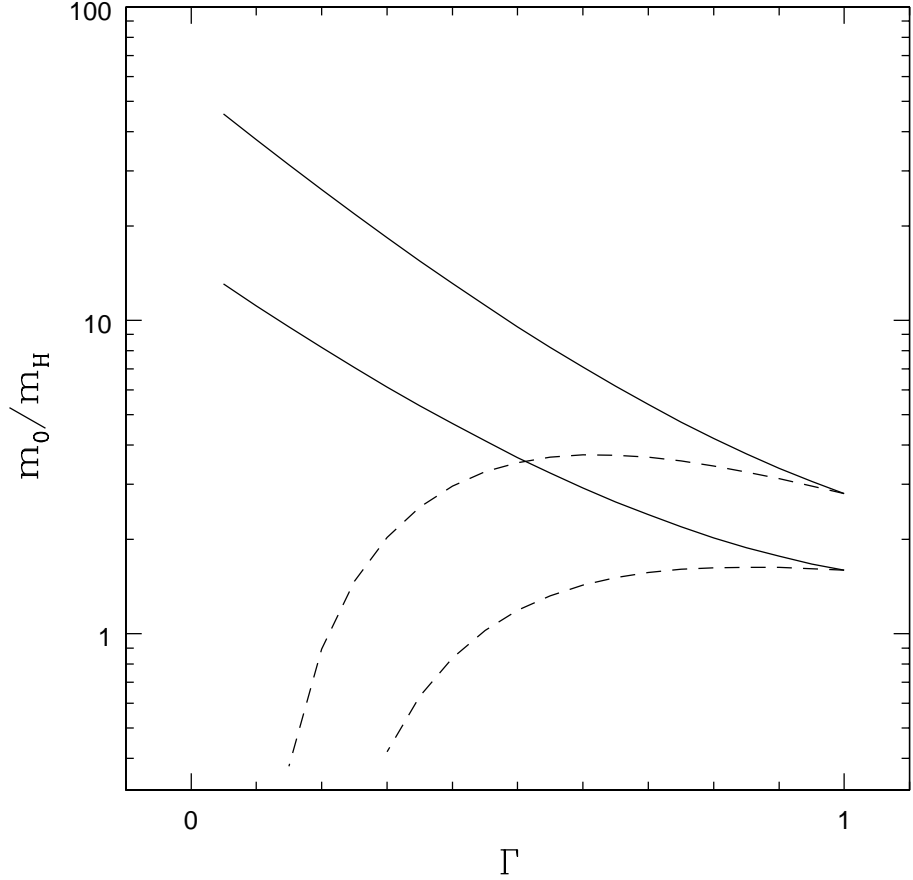} }}
\figcaption{Reduces mass ratio for varying static indices $\Gamma$ and
fixed (isothermal) dynamic $\gamma$ = 1. These curves show the
numerically determined values of $m_0/m_H$, which set the mass infall
rates, as a function of static index $\Gamma$ (the denominator uses
the hydrostatic value of $m_H$ with $\overdense$ = 1).  The solid
curves represent models in which the dynamic index is the same, i.e.,
$\gamma = \Gamma$. The dashed curves show cases where dynamic $\gamma
= 1$.  In other words, the polytropic spheres are allowed to have
varying initial density profiles (as set by varying $\Gamma$), but the
entropy evolution of the gas takes place under isothermal conditions
$\gamma = 1$. The upper curves correspond to initial states with
overdensity $\Lambda = 2$; the lower curves use overdensity $\Lambda =
1.5$. } \end{figure}

\newpage 
\begin{figure}
\figurenum{13}
{\centerline{\epsscale{0.90} \plotone{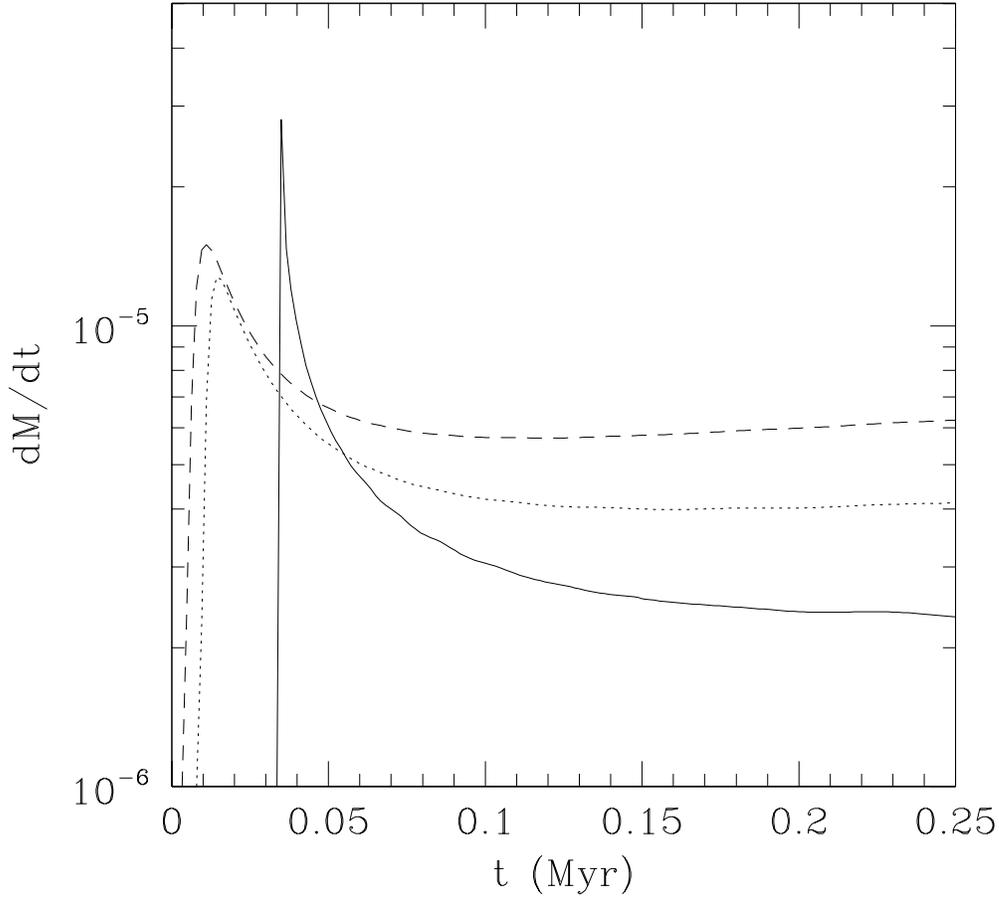} }} 
\figcaption{Mass infall rates for flat-topped cores as a function of
time for varying initial infall speeds. This set of numerical collapse
calculations begins with typical parameters observed in flat-topped
cores with sound speed $a_S$ = 0.2 km/s and a central region of
nearly constant density spanning $r_C \approx 10^{16}$ cm. The
initial states are overdense at the 10 percent level, so that $\abig$
= 2.2. The solid curve shows the resulting mass infall rate (expressed
in $M_\odot$ yr$^{-1}$) for the case with no initial velocity. The
dotted (dashed) curve shows the corresponding mass infall rate for
starting inward speeds of $\uin$ = 0.5 $a_S$ ($\uin$ = $a_S$). When 
$\uin \ne 0$, the mass infall rates reach their peak values more
quickly, the peak values are smaller, and the infall rates more
rapidly approach their asymptotic values (as predicted by the 
similarity solutions). }  
\end{figure}

\end{document}